\documentclass[pre,showkeys,preprintnumbers,amsmath,amssymb,superscriptaddress,twocolumn,nofootinbib]{revtex4}

\usepackage{graphicx}
\usepackage{amsfonts}
\usepackage{url}
\usepackage{color}
\usepackage{bm}
\usepackage[colorlinks=true,linkcolor=blue,citecolor=red]{hyperref}%
\usepackage{multirow}

\def\L{\mathcal L}
\def\N{\mathcal N}

\def\n{\bm{n}}
\def\s{\bm{s}}
\def\x{\bm{x}}

\def\X{\bm{X}}
\def\W{\bm{W}}

\def\k{\bm{k}}

\def\pa{{\partial\Omega}}
\def\E{{\mathbb E}}
\def\I{{\mathbb I}}
\def\P{{\mathbb P}}
\def\R{{\mathbb R}}

\def\T{{\mathcal T}}
\def\L{{\mathcal L}}
\def\F{{\mathcal F}}

\def\M{{\mathcal M}}

\def\erf{\mathrm{erf}}
\def\erfc{\mathrm{erfc}}
\def\erfcx{\mathrm{erfcx}}

\def\FF{\bm{F}}

\begin{document}

\title{Paradigm shift in diffusion-mediated surface phenomena}
\author{Denis~S.~Grebenkov}
 \email{denis.grebenkov@polytechnique.edu}
\affiliation{
Laboratoire de Physique de la Mati\`{e}re Condens\'{e}e (UMR 7643), \\ 
CNRS -- Ecole Polytechnique, IP Paris, 91128 Palaiseau, France}


\date{\today}

\begin{abstract}
Diffusion-mediated surface phenomena are crucial for human life and
industry, with examples ranging from oxygen capture by lung alveolar
surface to heterogeneous catalysis, gene regulation, membrane
permeation and filtration processes.  Their current description via
diffusion equations with mixed boundary conditions is limited to
simple surface reactions with infinite or constant reactivity.  In
this letter, we propose a probabilistic approach based on the concept
of boundary local time to investigate the intricate dynamics of
diffusing particles near a reactive surface.  Reformulating
surface-particle interactions in terms of stopping conditions, we
obtain in a unified way major diffusion-reaction characteristics such
as the propagator, the survival probability, the first-passage time
distribution, and the reaction rate.  This general formalism allows us
to describe new surface reaction mechanisms such as for instance
surface reactivity depending on the number of encounters with the
diffusing particle that can model the effects of catalyst fooling or
membrane degradation.  The disentanglement of the geometric structure
of the medium from surface reactivity opens far-reaching perspectives
for modeling, optimization and control of diffusion-mediated surface
phenomena.
\end{abstract}



\maketitle

\begin{figure}[!t]
\begin{center}
\includegraphics[width=80mm]{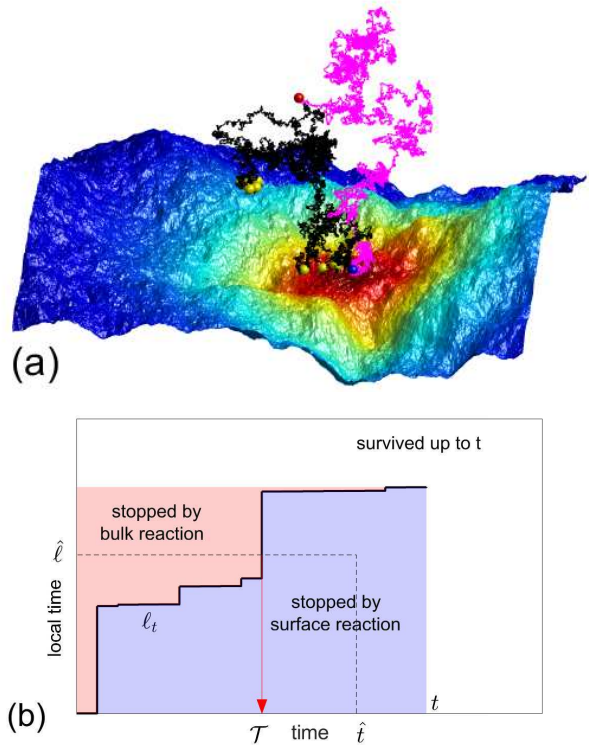}
\end{center}
\caption{
{\bf (a)} A simulated trajectory of a particle diffusing over a
reactive surface (see \ref{sec:Fig1} for details).  Red, blue and
yellow balls indicate respectively the starting bulk point $\x_0$, the
first arrival point $\s_0$ onto the surface, the consequent boundary
points at which the particle encountered the surface.  Pink and black
colors encode respectively the first segment of the trajectory (from
red to blue ball), and the remaining part.  {\bf (b)} The boundary
local time $\ell_t$ of the simulated trajectory (black thick line).
The statistics of $\ell_t$ is determined by reflected Brownian motion,
while bulk and surface reaction mechanisms are then incorporated by
introducing a stopping time $\hat{t}$ (vertical dashed line) and a
stopping local time $\hat{\ell}$ (horizontal dashed line),
respectively.  The surface reaction time $\T = \inf\{t > 0~:~ \ell_t >
\hat{\ell}\}$ is the random moment (indicated by arrow) when the
boundary local time $\ell_t$ crosses the horizontal line at
$\hat{\ell}$.}
\label{fig:surface}
%
\end{figure}

The dynamics of particles near a reactive surface is critically
important for many natural phenomena and industrial processes such as
diffusion-mediated heterogeneous catalysis, biochemical reactions on
DNA strands, proteins and cell membranes, filtration through porous
media, permeation across membranes, surface relaxation in nuclear
magnetic resonance, target searching and animal foraging, to name but
a few
\cite{Rice,Redner,Schuss,Metzler,Oshanin,Grebenkov07,Bressloff13,Benichou14}.
In a typical setting, a particle (e.g., a molecule, an ion, a protein,
a bacterium, an animal) moves inside a confining medium; when the
particle comes close to the boundary of the medium, an appropriate
surface mechanism can be initiated, e.g., the particle can bind to the
boundary, relax its fluorescence, magnetization or another form of
excitation, be chemically transformed into another molecule, be
transported through a membrane pore, or be killed or destroyed (all
these distinct mechanisms will be generically called ``surface
reaction'' in the following).  Whatever the surface mechanism is, its
successful realization is not granted and depends on the state of the
local environment near the particle.  For instance, the boundary can
be locally inert for binding, possess no catalytic germ or impurity
for chemical transformation or relaxation; the closest membrane pore,
channel or gate can be temporarily inactive or already occupied, while
a predator can be asleep or not hungry; even if the target molecule or
the escape hole is found, the particle may not overcome an energy
activation or entropic barrier.  In any of such unfavorable
circumstances, the particle resumes its bulk motion until the next
arrival to the boundary, and so on \cite{Bychuk95,Wang17}.  As a
consequence, the successful realization of the surface reaction is
typically preceded by a long sequence of successive bulk explorations,
started and terminated on the surface (Fig. \ref{fig:surface}(a)).
Even for ordinary bulk diffusion, partial surface reactivity results
in very intricate and still poorly understood dynamics that affects
the functioning of chemical reactors, living cells, exchange devices
and organs such as lungs and placenta
\cite{Rice,Redner,Schuss,Metzler,Oshanin,Grebenkov07,Bressloff13,Benichou14,Bychuk95,Wang17,Sapoval02,Grebenkov05,Serov16}. 
This dynamics becomes even more sophisticated for mortal walkers
\cite{Yuste13,Meerson15,Grebenkov17d} that have a finite random
lifetime due to, for instance, bulk relaxation, photobleaching,
radioactive decay, bulk reaction, or starving.

The conventional description of these phenomena relies on macroscopic
concentrations or, more fundamentally, on a propagator (also known as
heat kernel or Green's function), $G_q(\x,t|\x_0)$, that characterizes
the likelihood of finding a particle that started from a point $\x_0$
at time $0$ and {\it survived} (not reacted) up to time $t$, in a bulk
point $\x$ at time $t$.  The propagator obeys the Fokker-Planck
equation, in which the bulk dynamics dictates the form of the
Fokker-Planck operator, whereas the shape and the reactivity of the
surface set boundary conditions \cite{Risken}.  For ordinary bulk
diffusion, Collins and Kimball \cite{Collins49} put forward the Robin
(also known as Fourier, radiation, or third) boundary condition
\begin{equation}  \label{eq:Robin}
- D \, \partial_n G_q(\x,t|\x_0) = \kappa \, G_q(\x,t|\x_0), 
\end{equation}
where $D$ is the diffusion coefficient and $\partial_n$ is the normal
derivative.  At each boundary point, the net diffusive flux density
from the bulk (the left-hand side) is equated to the reaction flux
density, which is {\it postulated} to be proportional to $G_q(\x,t|\x_0)$
on the boundary.  The proportionality coefficient $\kappa$ (in units
of speed, m/s) bears the names of reactivity, permeability,
relaxivity, or inverse surface resistance
\cite{Sapoval94,Grebenkov06a}, and can be related to the on-rate
constant $k_{\rm on}$ of chemical reactions
\cite{Lauffenburger,Sano79,Shoup82},
to the microscopic heterogeneity of catalytic germs
\cite{Zwanzig90,Berezhkovskii04,Bernoff18b,Grebenkov19},
to the opening dynamics of gates, channels or pores
\cite{Benichou00,Reingruber09}, to the energy activation or entropic
barrier \cite{Grebenkov17}, and to the probability of the reaction
event at each encounter \cite{Grebenkov03,Grebenkov06}.  The interplay
between diffusive transport from the bulk and reaction on the surface
is controlled by the ratio $q = \kappa/D$, ranging from $0$ for an
inert boundary, to infinity for a perfectly reactive boundary.  The
inverse of $q$, $1/q$, sets a characteristic reaction length
\cite{Sapoval94,Sapoval02,Grebenkov03,Grebenkov06a}.  As the surface
reaction mechanism is incorporated via the boundary condition
(\ref{eq:Robin}), the dependence of the propagator $G_q(\x,t|\x_0)$ on
the reactivity $\kappa$ (or $q$) is {\it implicit} that impedes
studying these phenomena and optimizing shapes and reactivity patterns
of catalysts or clustering of receptors/pores on the cell membrane.

In this letter, we advocate for an alternative description of
diffusion-mediated surface phenomena based on the concept of boundary
local time.  The main text describes our findings in a general and
broadly accessible but still rigorous way (with a limited number of
formulas), whereas the Supplemental Material (SM) provides all the
necessary details for theoreticians.
We recall that reflected Brownian motion $\X_t$ in a confining domain
$\Omega \subset \R^d$ with a smooth {\it inert} boundary $\pa$ is
mathematically constructed as the solution of the stochastic Skorokhod
equation \cite{Ito,Freidlin,Saisho87}:
\begin{equation}  \label{eq:Skorokhod0} 
d\X_t = \sqrt{2D} \, d\W_t + \n(\X_t) \, d\ell_t,  \quad \X_0 = \x_0,
\end{equation}
where $\W_t$ is the standard Brownian motion, $\n(\x)$ is the unit
normal vector, and $\ell_t$ (with $\ell_0 = 0$) is a nondecreasing
process, which increases only when $\X_t\in\pa$, known as the boundary
local time (see Sec. \ref{sec:local_time} of the SM for a discussion of
this concept).  Qualitatively, Eq. (\ref{eq:Skorokhod0}) can be
understood as a Langevin equation with a very strong short-range
repulsive force localized on the boundary.  Indeed, the second term in
Eq. (\ref{eq:Skorokhod0}) is nonzero only for $\X_t \in
\pa$ and ensures that the particle is reflected in the perpendicular
direction $\n(\x)$ from the boundary at each encounter.  The peculiar
feature of this construction is that the single Skorokhod equation
determines simultaneously two tightly related stochastic processes:
$\X_t$ and $\ell_t$.  The conventional propagator $G_0(\x,t|\x_0)$
(with $q = 0$) characterizes the position $\X_t$ of the diffusing
particle but ignores its boundary local time $\ell_t$.  But it is
precisely the local time that bears information on particle's
encounters with the boundary and is thus the key ingredient to account
for surface reactions.  We therefore build an alternative description
on the full propagator $P(\x,\ell,t|\x_0)$, i.e., the joint
probability density of both $\X_t$ and $\ell_t$ at time $t$.  Due to
the jump-like character of the boundary local time (see
Fig. \ref{fig:surface}(b)), finding the full propagator was the most
challenging and mathematically involved part of this work (see
Sec. \ref{sec:alternative}).

As the full propagator characterizes the diffusive dynamics alone
(without reactions), it is the most natural theoretical ground, to
which both bulk and surface reactions can be added {\it explicitly}
via stopping conditions.  Indeed, if the diffusing particle can
spontaneously die, relax its excitation, be chemically transformed,
killed or destroyed in the bulk, its finite lifetime can be modeled by
a random stopping time $\hat{t}$.  In a common Poissonian setting,
such a bulk reaction can occur at any time instance with equal chances
(characterized by rate $p$), so that the lifetime of the particle
obeys the exponential distribution: $\P\{ \hat{t} > t\} = e^{-pt}$.
If this bulk reaction mechanism is independent of the diffusive
dynamics, averaging the full propagator $P(\x,\ell,\hat{t}|\x_0)$ over
random realizations of $\hat{t}$ yields the joint distribution of
$\X_{\hat{t}}$ and $\ell_{\hat{t}}$ at the moment $\hat{t}$ of bulk
reaction (or particle's death).  More generally, one can introduce
elaborate stopping times to incorporate eventual delays in the bulk
diffusion due to reversible binding to immobile centers or mobile
buffer molecules (like waiting time distribution in continuous-time
random walks), time-dependent or switching diffusivity, the effects of
rapidly re-arranging dynamic medium, and other subordination
mechanisms (see \cite{Metzler00,Chechkin17,Lanoiselee18} and
references therein).

Remarkably, we discovered that surface reaction mechanisms can be
implemented in essentially the same way.  At each encounter with the
partially reactive surface, the particle either reacts with the
probability $\Pi = 1/(1 + D/(\kappa a)) \simeq aq \ll 1$, or resumes
its bulk diffusion with the probability $1-\Pi$, where $a$ is the
width of a thin reactive boundary layer (i.e., the interaction range)
\cite{Grebenkov03}.  If all reaction attempts are independent from
each other, the number $\hat{n}$ of failed attempts until the
successful reaction follows the geometric law, $\P\{ \hat{n} > n\} =
(1 - \Pi)^n \approx \exp(-q na)$.  The rescaled number of failed
reaction attempts, $\hat{\ell} = a \hat{n}$, obeys thus the
exponential law: $\P\{ \hat{\ell} > \ell \} = e^{-q\ell}$, with $\ell
= an$.
As the boundary local time $\ell_t$ is related to the number
$\N_t^{a}$ of encounters of the particle with the surface up to time
$t$, $\ell_t = \lim\limits_{a\to 0} a \,\N_t^{a}$ \cite{Ito,Freidlin}
(see also \ref{sec:local_time}), the reaction time $\T$ can be defined as
the moment, at which $\ell_t$ exceeds an independent random stopping
local time $\hat{\ell}$: $\T = \inf\{ t> 0 ~:~ \ell_t > \hat{\ell}\}$
\cite{Grebenkov06,Grebenkov07a,Grebenkov19c}.  Multiplication of the full
propagator $P(\x,\ell,t|\x_0)$ by the probability $\P\{ \hat{\ell} >
\ell \} = e^{-q\ell}$ of no surface reaction up to $\ell$ and
integration over $\ell$ yield the marginal propagator of the position
$\X_t$ at time $t$ of a particle, conditioned to survive up to time
$t$ (i.e., with the condition $\T > t$, which is equivalent to
$\hat{\ell} > \ell_t$).  By construction, this average is precisely
the conventional propagator:
\begin{equation}  \label{eq:Pfull_Gq}
G_q(\x,t|\x_0) = \int\limits_0^\infty d\ell \, e^{-q\ell} \, P(\x,\ell,t|\x_0).
\end{equation}
To our knowledge, this fundamental relation was not reported earlier.

Moreover, changing the distribution of the stopping local time
$\hat{\ell}$, one can now easily implement new surface reaction
mechanisms.  In fact, the average of the full propagator
$P(\x,\ell,t|\x_0)$ with the probability of no surface reaction up to
$\ell$, now determined by a desired distribution $\Psi(\ell) =
\P\{\hat{\ell} > \ell\}$ of $\hat{\ell}$, yields a generalized
propagator
\begin{equation}
G_\psi(\x,t|\x_0) = \int\limits_0^\infty d\ell \, \Psi(\ell) \, P(\x,\ell,t|\x_0).
\end{equation}
This relation couples explicitly the surface reaction mechanism
(represented by $\Psi(\ell)$) and the dynamics of the particle
diffusing in a domain with reflecting boundary (represented by
$P(\x,\ell,t|\x_0)$).  The striking similarity of our implementations
of bulk and surface reactions is not surprising: while time $t$ mimics
the number of bulk steps (and thus exposure of the particle to bulk
reaction), the local time $\ell$ counts the number of encounters with
the boundary (and its exposure to surface reaction).  It is crucial
that both bulk and surface reaction mechanisms, introduced via two
independent random variables $\hat{t}$ and $\hat{\ell}$
(Fig. \ref{fig:surface}(b)), are disentangled from the dynamics.  In
other words, one can first investigate the dynamics in the case of
reflecting boundary and then couple it explicitly to reaction
mechanisms.

The alternative description allows one to go far beyond the constant
reactivity based on the Robin boundary condition (\ref{eq:Robin}).
Indeed, the former exponential law $\Psi(\ell) = e^{-q\ell}$ described
a Poissonian-like mechanism when the particle could react at each
encounter with the boundary with equal probabilities.  To incorporate
variable reaction probabilities, we introduce the reactivity
$\kappa(\ell)$ that changes with the local time $\ell$ (i.e., with the
rescaled number of encounters), alike time-dependent diffusivity
$D(t)$ for bulk diffusion.  Extending our previous arguments (see
\ref{sec:passivation}), we derive the probability distribution for the
corresponding stopping local time $\hat{\ell}$:
\begin{equation}  \label{eq:psi_kappa0}
\Psi(\ell) = \exp\biggl( - \frac{1}{D} \int\limits_0^\ell d\ell' \, \kappa(\ell') \biggr).
\end{equation}
This is a new feature brought by our probabilistic description, which
allows us to investigate within the unique theoretical framework many
important diffusion-mediated surface phenomena such as catalyst's
fooling or membrane degradation \cite{Bartholomew01,Filoche08}.  In
fact, choosing an appropriate $\kappa(\ell)$ (or $\Psi(\ell)$), one
can control the reaction dynamics of the boundary.  For instance, the
reactivity $\kappa(\ell)$, which is small at $\ell\approx 0$ and then
reaches a constant level, can model situations when the surface needs
to be progressively activated by repeated encounters with the
diffusing particle.  In contrast, when $\kappa(\ell)$ is large at
small $\ell$ and then reaches a constant (or vanishes), one models a
progressive passivation of initially highly reactive surfaces.

The generalized propagator $G_\psi(\x,t|\x_0)$ determines other common
characteristics of diffusion-reaction processes such as, e.g., the
survival probability or the reaction rate (see Fig.~\ref{fig:scheme} and
\ref{sec:generalized}).  For instance, we show in the SM that the
probability density of the first-passage (or reaction) time $\T$ can
be written as
\begin{equation}  \label{eq:H_U}
H_\psi(t|\x_0) = \int\limits_0^\infty d\ell \, \psi(\ell) \, U(\ell,t|\x_0),
\end{equation}
where $\psi(\ell) = - d\Psi(\ell)/d\ell$ is the probability density of
the stopping local time $\hat{\ell}$, and $U(\ell,t|\x_0) = D
\int\nolimits_{\pa} d\s \, P(\s,\ell,t|\x_0)$ is the probability
density of the first-crossing time of a level $\ell$ by the boundary
local time $\ell_t$ (see \ref{sec:SI_U}).  This relation expresses the
idea illustrated in Fig. \ref{fig:surface}(b): the surface reaction
occurs when the boundary local time $\ell_t$ exceeds a random level
$\hat{\ell}$ determined by $\psi(\ell)$.
For a perfectly reactive boundary, the reaction occurs at the first
encounter with the boundary, i.e., at the first moment when the
boundary local time exceeds $0$.  This is precisely the first-crossing
time of the level $0$, i.e., $\psi(\ell) = \delta(\ell)$, and thus
$U(0,t|\x_0)$ is the probability density of the common first-passage
time to a perfect target \cite{Benichou10b}.  In turn,
$U(\ell,t|\x_0)$ for any $\ell > 0$ describes the reaction time in the
case when the reaction occurs at the boundary local time $\ell_t =
\ell$ (i.e., after a prescribed number of failed reaction attempts).
According to Eq. (\ref{eq:H_U}), other surface reaction mechanisms can
be described by setting the level $\ell$ randomly, i.e., by
introducing the stopping local time $\hat{\ell}$.  

\begin{figure}[!t]
\begin{center}
\includegraphics[width=88mm]{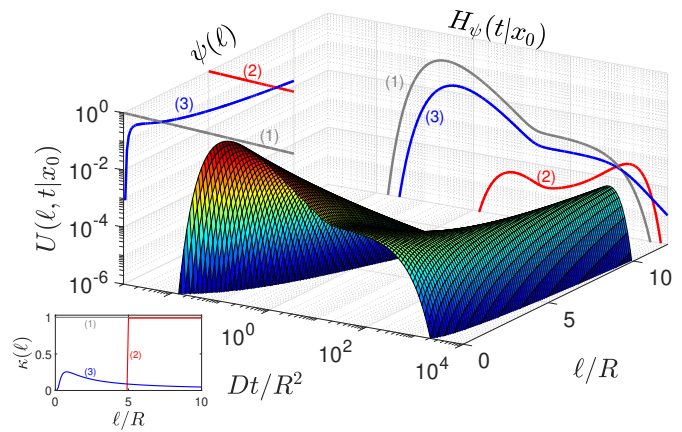}
\end{center}
\caption{
The probability density $U(\ell,t|\x_0)$ (rescaled by $R^2/D$) of the
first-crossing time of a level $\ell$ by the boundary local time
$\ell_t$ for a spherical target of radius $R$, surrounded by an outer
reflecting concentric sphere of radius $L$, with $|\x_0|/R = 2$ and
$L/R = 10$.  Three curves on the left projection illustrate three
probability densities $\psi(\ell)$ of the stopping local time
$\hat{\ell}$ (with $qR = 1$): (1) $\psi(\ell) = qe^{-q\ell}$ (gray
line, conventional setting); (2) $\psi(\ell) = qe^{-q\ell}
\Theta(\ell-\ell_0)$ with $\ell_0/R = 5$ (red line);
and (3) $\psi(\ell) = q e^{-1/(q\ell)} /[\sqrt{\pi} (q\ell)^{3/2}]$
(blue line).  These densities correspond respectively to three
reactivity profiles shown on the inset: $\kappa(\ell) = qD$,
$\kappa(\ell) = qD \Theta(\ell - \ell_0)$, and $\kappa(\ell) = qD
e^{-1/(q\ell)} /[\sqrt{\pi} (q\ell)^{3/2} \erf(1/\sqrt{q\ell})]$.
Three curves on the right projection show the corresponding
probability densities $H_\psi(t|\x_0)$ from Eq. (\ref{eq:H_U}). }
\label{fig:U_shell}
\end{figure}

Figure \ref{fig:U_shell} exemplifies the impact of surface reaction
mechanisms onto the distribution of the reaction time.  Here the
family of the probability densities $U(\ell,t|\x_0)$ (parameterized by
$\ell$) is presented for a spherical target, surrounded by an outer
reflecting sphere.  Three probability densities $\psi(\ell)$
determining surface reaction mechanisms (with $\kappa(\ell)$ in the
inset) are plotted on the left projection, while the resulting
reaction time densities $H_\psi(t|\x_0)$ are shown on the right
projection.  For a constant reactivity, the average of
$U(\ell,t|\x_0)$ with the exponential density $qe^{-q\ell}$ (gray
line) results in the conventional reaction time distribution
\cite{Grebenkov18}.  Here, a single ``hump'' region around the most
probable reaction time is followed by a flat part and ultimate
exponential decay, as it should be for a bounded domain.  If the
target is passive at the beginning (red line), first arrivals of the
particle onto the target do not produce reaction up to some local time
$\ell_0$, thus shifting the probability density of the reaction time
to longer times.  Curiously, an unusual second ``hump'' region emerges
due to the particles that moved away from the target, explored the
confining domain and then returned to the target (see also
\ref{sec:surf_mech}).  In the third example (blue line), the reactivity is
negligible at the beginning, reaches a maximum around $\ell/R \approx
0.7$, and then slowly decreases as $1/(2q\ell)$ at large $\ell$.  The
overall shape of the reaction time density $H_\psi(t|\x_0)$ resembles
that of the conventional setting, but exhibits anomalous power law
decay at long times: $H_\psi(t|\x_0) \propto t^{-3/2}$.  Here, as the
encounter-dependent reactivity offers an optimal range of local times
for surface reaction, a particle that failed to react during this
range, has lower and lower chances to react after more or more returns
to the target.

More generally, the asymptotic large-$\ell$ decay $\kappa(\ell)
\propto 1/\ell$ turns out to be the critical regime that distinguished
three scenarios for arbitrary bounded domains (see
\ref{sec:surf_mech}): (i) if $\kappa(\ell)$ decays slower than $1/\ell$
(or increases with $\ell$), $H_\psi(t|\x_0)$ exhibits the long-time
exponential decay, as in the conventional setting; (ii) if
$\kappa(\ell)$ decreases as $\nu D/\ell$ with some constant $0 < \nu <
1$, then $H_\psi(t|\x_0) \propto t^{-1-\nu}$; this is a new unexpected
feature for bounded domains; (iii) if $\kappa(\ell)$ decays faster
than $1/\ell$, the reaction time can be infinite with a finite
probability:
\begin{equation}
\P\{ \T = \infty\} = \Psi(\infty) = \exp\biggl(-\int\limits_0^\infty d\ell \, \frac{\kappa(\ell)}{D} \biggr) > 0.
\end{equation}
In other words, this is the probability of no surface reaction in a
bounded domain: even though the exploration is compact, the reactivity
decays too fast so that the particle may fail to react even after an
infinite number of returns to the target.  Several surface reaction
models and the behavior of the underlying reaction time distributions
and reaction rates are discussed in \ref{sec:surf_mech}.

In summary, we developed a new probabilistic description of
diffusion-mediated surface phenomena based on the concept of boundary
local time.  By introducing the full propagator to describe confined
diffusion with reflections on the boundary, we succeeded to
incorporate surface reactivity {\it explicitly} via a stopping
condition.  The disentanglement of the surface reactivity from the
dynamics allowed us to introduce encounter-dependent reactivity
$\kappa(\ell)$ and to describe a variety of new surface reaction
mechanisms.  We discussed how different forms of $\kappa(\ell)$ affect
the reaction times and revealed some intriguing anomalous features in
their distribution.

The developed formalism opens a vast area for future research.  On the
theoretical side, one can study how the diffusive dynamics in domains
with complex geometric structures (such as the interior of an
eukaryotic cell, a chemical reactor, or a human acinus) is coupled to
different surface reaction mechanisms.  One can further extend this
approach to investigate (i) the (anti-)cooperativity effects of
multiple diffusing particles whose individual encounters with the
boundary change its reactivity; (ii) the competition between multiple
targets, each described by its own boundary local time; (iii) the
combined impact of bulk and surface reaction mechanisms; (iv) the
effects of correlations between successive encounters, and (v)
the presence of long-range interactions with and reversible binding to
the boundary.  In particular, our probabilistic description of the
bulk exploration step until reversible binding to the boundary can
bring complementary insights to former theoretical approaches based on
coupled diffusion-reaction equations
\cite{Chechkin09,Berezhkovskii15} (see \ref{sec:Chechkin}).
On the application side, appropriate surface reaction models should be
identified to describe industrial examples of catalyst fooling,
membrane aging and many other diffusion-mediated surface phenomena, in
which the surface properties depend on the number of encounters.  One
can also address a new class of optimization problems targeting
optimal reaction rates or prescribed distributions of reaction times
or positions, either by adapting the surface reaction mechanisms for a
given geometric structure of the medium, or by optimizing its
structure for a given surface reaction mechanism, or both.  The
disentanglement of the geometric structure from the surface reaction
mechanism is the key that has now opened the door to such
applications.

\begingroup
\renewcommand{\addcontentsline}[3]{}
\renewcommand{\section}[2]{}

\endgroup

\pagebreak ~
\pagebreak

\begin{center}
\textbf{SUPPLEMENTAL MATERIAL}
\end{center}
\label{SM}
\setcounter{equation}{0}
\setcounter{section}{0}
\setcounter{figure}{0}
\setcounter{table}{0}
\setcounter{footnote}{0}
\makeatletter
\renewcommand{\theequation}{S\arabic{equation}}
\renewcommand{\thefigure}{S\arabic{figure}}
\renewcommand{\thesection}{SM.\Roman{section}}

\tableofcontents

\section{Boundary local time}
\label{sec:local_time}

While the boundary local time is a cornerstone of the mathematical
theory of stochastic processes \cite{Ito,Freidlin}, this important
classic concept remains largely unknown in physical, chemical and
biological communities.  As the boundary local time plays the central
role in our probabilistic description of diffusion-mediated surface
phenomena, we provide here a short introduction for physicists.

The notion of a (point) local time has been first introduced by L\'evy
to quantify a fraction of time that Brownian motion spent at that
point \cite{Levy}.  The properties of point local times were
thoroughly investigated, in particular, for Brownian motion and Bessel
processes (see \cite{Borodin} and references therein).
This notion was extended to boundaries and applied for a rigorous
mathematical construction of diffusive processes that are confined
inside a given domain $\Omega$ with a smooth boundary $\pa$
\cite{Ito,Freidlin}.  The distribution of the boundary local time for
some Euclidean domains was investigated
\cite{Grebenkov07a,Grebenkov19c} (see also
\cite{Hsu85,Williams87,Takacs95,Zhou17}) and will also be discussed in
Sec. \ref{sec:BLT}.

To get an intuitive feeling of this construction, let us consider the
Langevin equation for a particle diffusing in a force field $\FF(\x)$:
\begin{equation}
d\X_t = \frac{D}{k_B T}  \FF(\X_t) \, dt + \sqrt{2D} \, d\W_t ,
\end{equation} 
where $\W_t$ is the standard Brownian motion (representing the
thermally induced motion), $D$ is the diffusion coefficient, and
$D/(k_B T)$ is the drag coefficient related to the fluid viscosity
(with $k_B$ being the Boltzmann constant and $T$ the absolute
temperature).  As Brownian motion $\W_t$ can explore the whole
Euclidean space, the force field is needed to keep the particle inside
the domain $\Omega$.  If one aims at constructing reflected Brownian
motion (i.e., ordinary diffusion with reflections on $\pa$), the force
field should not affect the motion inside the domain.  In other words,
$\FF(\x)$ should be zero inside $\Omega$, except for a very thin
boundary layer of width $a$, in which $\FF(\x)$ should force the
particle to move away from the boundary back to the interior (a sort
of repulsion).  Moreover, as the force acts only in that thin layer,
its repulsive action should be strong enough to avoid any crossing of
the boundary.  A simple choice of the force field is
\begin{equation}
\frac{\FF(\x)}{k_B T} = \frac{1}{a} \, \n(\x) \, \I_{\pa_a}(\x), 
\end{equation}
where $\n(\x)$ is the normal vector to the boundary $\pa$ determining
the force direction back to the bulk in the orthogonal direction (to
avoid any longitudinal bias along the boundary), $\pa_a = \{
\x\in\Omega ~:~ |\x - \pa| < a\}$ is the boundary layer of width $a$,
inside which the force acts, and $\I_{\pa_a}(\x)$ is the indicator
function of $\pa_a$: $\I_{\pa_a}(\x) = 1$ if $\x\in \pa_a$ and $0$
otherwise.  The factor $1/a$ increases the force amplitude when the
layer width diminishes.  Denoting
\begin{equation} \label{eq:ell_ta}
\ell_t^a = \frac{D}{a} \int\limits_0^t dt' \,  \I_{\pa_a}(\X_{t'}),
\end{equation}
the Langevin equation takes the form
\begin{equation}
d\X_t = \n(\X_t) \, d\ell_t^a + \sqrt{2D} \, d\W_t .
\end{equation} 
By definition, $(a/D) \ell_t^a$ is the residence (or occupation) time
of the process $\X_t$ inside the boundary layer $\pa_a$ up to time
$t$.  As the boundary layer is getting thinner ($a\to 0$), its volume
and thus the residence time in $\pa_a$ vanish.  However, the rescaling
by $1/a$ ensures a nontrivial limit,
\begin{equation}  \label{eq:ell_res}
\ell_t = \lim\limits_{a\to 0} \ell_t^a ,
\end{equation}
while the above Langevin equation yields the stochastic Skorokhod
equation
\begin{equation}  \label{eq:Skorokhod}
d\X_t = \n(\X_t) \, d\ell_t + \sqrt{2D} \, d\W_t .
\end{equation} 
The stochastic process $\ell_t$ is called the boundary local time.
Even though $\ell_t$ has units of length (due to the factor $D/a$ in
Eq. (\ref{eq:ell_ta})), it can still be thought of as a fraction of
time that reflected Brownian motion spent in an infinitesimal vicinity
of the boundary up to time $t$.  By construction, the first term in
Eq. (\ref{eq:Skorokhod}) is nonzero only when $\X_t$ is on the
boundary (to highlight this property, the indicator function
$\I_{\pa}(\X_t)$ is sometimes included explicitly to the first term).
We hasten to stress that this is not a rigorous derivation of the
Skorokhod equation but rather its intuitive physical explanation.

\begin{figure}
\begin{center}
\includegraphics[width=88mm]{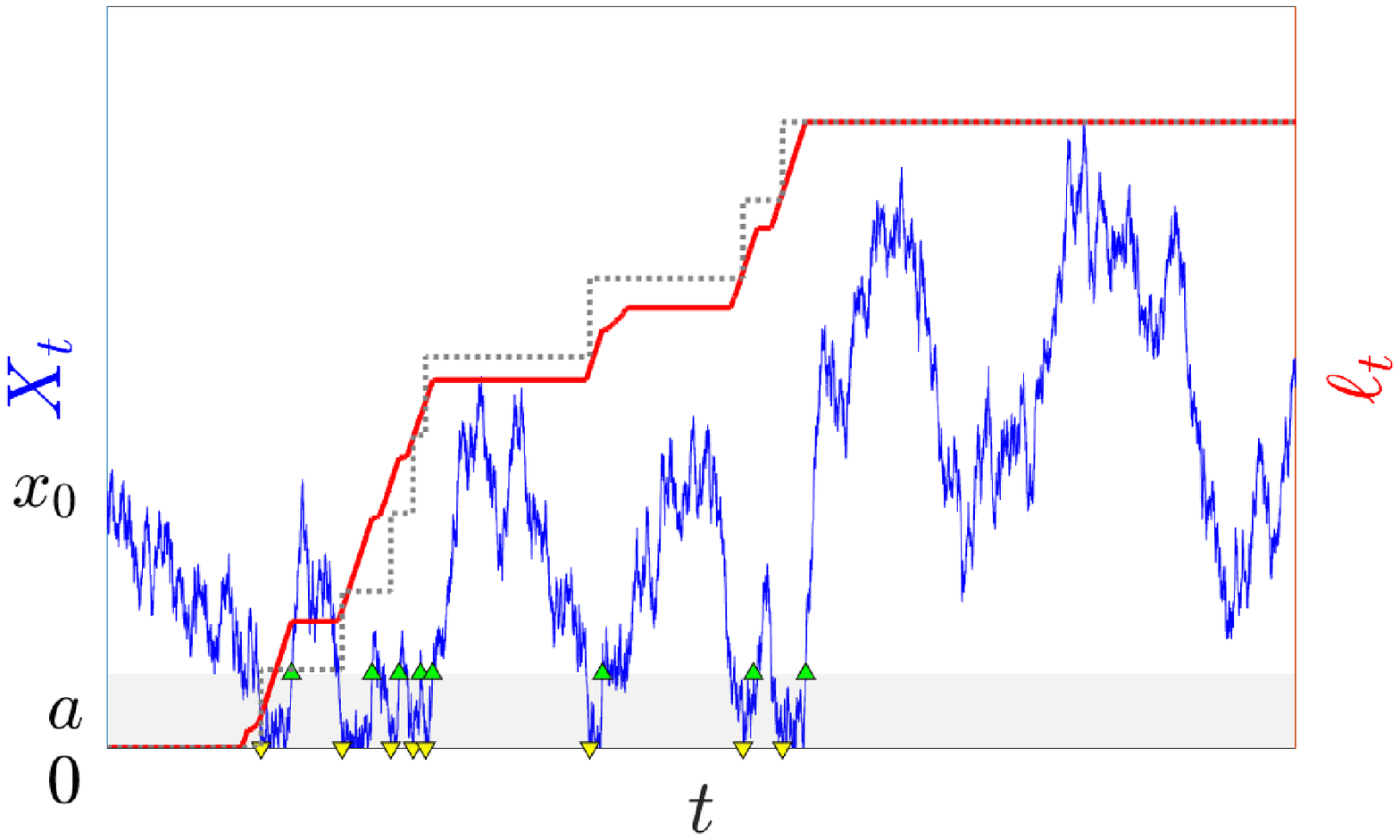}
\end{center}
\caption{
Simulated random trajectory (blue line) of reflected Brownian motion
$X_t$ on the half-line $\Omega = (0,\infty)$ (vertical axis) and two
approximations $\ell_t^a$ (red solid line) and $a\,\N_t^a$ (gray
dotted line) of the boundary local time $\ell_t$ at the origin, all
plotted as functions of time $t$.  Shadowed region outlines the zone
when the particle diffuses inside the boundary layer $\pa_a = (0,a)$
near the origin ($\pa = \{0\}$).  Yellow down-pointing triangles
indicate the hitting times $\delta_n^{(0)}$ when the particle
downcrossed the boundary layer; in turn, green up-pointing triangles
show the hitting times $\delta_n^{(a)}$ when the particle left the
boundary layer after crossing.  Note that $X_t$, $\ell_t^a$ and $a\,
\N_t^a$ are rescaled to the same maximum for easier visualization.}
\label{fig:Xt_1d}
\end{figure}

The boundary local time $\ell_t$ can also be related to the number
$\N_t^{a}$ of (down)crossings of the boundary layer $\pa_a$ by
reflected Brownian motion up to time $t$.  This number can be defined
by introducing a sequence of interlacing hitting times $0 \leq
\delta_1^{(0)} < \delta_1^{(a)} < \delta_2^{(0)} < \delta_2^{(a)} <
\ldots$ as
\begin{subequations}
\begin{align}
\delta_n^{(0)} &= \inf\{ t > \delta_{n-1}^{(a)} ~:  ~\X_t \in \pa\},  \\
\delta_n^{(a)} &= \inf\{ t > \delta_n^{(0)} ~:  ~\X_t \in \Gamma_a\},
\end{align}
\end{subequations}
(with $\delta_{0}^{(0)} = \delta_{0}^{(a)} = 0$), where $\Gamma_a = \{
\x\in\Omega ~:~ |\x - \pa| = a\}$ (see, e.g., \cite{Grebenkov19c}).
Here, one records the first moment $\delta_1^{(0)}$ when reflected
Brownian motion hits the boundary $\pa$, then the first moment
$\delta_1^{(a)}$ of leaving the thin layer $\pa_a$ through its inner
boundary $\Gamma_a$, then the next moment $\delta_2^{(0)}$ of hitting
the boundary $\pa$, and so on.  The number of downcrossings of the
thin layer $\pa_a$ up to time $t$ is then the index $n$ of the largest
hitting time $\delta_n^{(0)}$, which is below $t$:
\begin{equation}
\N_t^{a} = \sup\{ n \geq 0 ~:~ \delta_n^{(0)} < t\} .
\end{equation}
While the number of downcrossings diverges as $a\to 0$, its rescaling
by $a$ yields the boundary local time \cite{Ito,Freidlin}:
\begin{equation} \label{eq:ell_def} 
\ell_t = \lim\limits_{a \to 0} a \, \N_t^{a} .
\end{equation}
As the number of encounters of the process $\X_t$ with the boundary
layer $\pa_a$ up to time $t$ can be naturally identified with the
number $\N_t^a$ of its downcrossings, the boundary local time divided
by the layer width, $\ell_t/a$, is a proxy of the number of
encounters, as soon as $a$ is small enough.  Note that the
approximations $a\, \N_t^a$ and $\ell_t^a$ are closely related.
Indeed, the residence time in Eq. (\ref{eq:ell_ta}) can be split into
separate contributions associated to each downcrossing, and each
contribution is of the order $a^2/D$ (the average time spent by
reflected Brownian motion in a thin boundary layer $\pa_a$).

By construction, the boundary local time $\ell_t$ is a nondecreasing
stochastic process, which remains $0$ until the first encounter with
the boundary.  After that, $\ell_t$ increases by tiny (infinitely
small) jumps at every encounter with the boundary.  These increments
can also be interpreted as increases of the residence time spent near
the boundary.
In turn, the time interval between two successive jumps in $\ell_t$
can be either small, or large.  In fact, reflected Brownian motion
hitting a smooth surface is known to return infinitely many times to
that surface within an infinitely short time period \cite{Morters}.
Even if each of these returns gives a tiny increment to the boundary
local time, their huge number results in notable changes of $\ell_t$.
In contrast, when the particle diffuses inside the domain, the
boundary local time remains constant until the next hit.

Figure \ref{fig:Xt_1d} illustrates two approximations $\ell_t^a$ and
$a\, \N_t^a$ of this process for reflected Brownian motion on a
half-line.  One can see that $\ell_t^a$ increases gradually due to its
integral form (\ref{eq:ell_ta}), whereas $a\,\N_t^a$ changes by jumps
of height $a$ at each downcrossing of the boundary layer.  While these
approximations behave quite differently for any finite $a$, both of
them converge to the boundary local time $\ell_t$ as $a\to 0$.  

We conclude this section by highlighting some features of reflected
Brownian motion $\X_t$ and its boundary local time $\ell_t$.  The
contact of the diffusing particle with the reflecting boundary is
instantaneous, the particle does not stay on that boundary as there is
no binding.  However, the self-similar nature of Brownian motion
results in infinitely many returns of this process to the boundary
whose accumulation yields nonzero increments in $\ell_t$.  We stress
that there is no attractive force to keep the particle close to the
boundary, it is just a probabilistic consequence of white noise
thermal fluctuations.  If some of these properties may sound
unphysical, it is because the mathematical processes $\X_t$ and
$\ell_t$ are the limits of their ``regularized'', more physically
appealing versions (e.g., a random walk $\X_t^a$ for $\X_t$ and
$\ell_t^a$ or $a\, \N_t^a$ for $\ell_t$) when $a\to 0$.  In a physical
system, there is always a natural boundary layer (set, e.g., by atomic
short-range interactions) so that one can keep thinking of $\X_t$ and
$\ell_t$ in terms of $\X_t^a$ and $\ell_t^a$ (or $a\, \N_t^a$).
Likewise, Monte Carlo techniques can only simulate ``regularized''
processes.  However, the crucial advantage of the limiting processes
$\X_t$ and $\ell_t$ is that they do not depend on the boundary layer
width $a$, and their properties are thus more universal and in general
easier to investigate by mathematical tools.  Once these properties
are established, one can use them to characterize systems with a small
finite $a$.

\section{Derivation of main results}
\label{sec:alternative}

In this section, we derive the main results of the letter.  While this
derivation employs and extends the arguments earlier developed
\cite{Grebenkov19,Grebenkov19b,Grebenkov19c}, the probabilistic
approach proposed in the letter and its major elements such as the
full propagator, the generalized diffusion-reaction characteristics,
and the encounter-dependent reactivity, are new.
We hasten to stress that the following presentation does not provide
mathematical proofs, so that its rigorous formulation and extensions
present interesting perspectives for future work.

\begin{figure}[!t]
\begin{center}
\includegraphics[width=88mm]{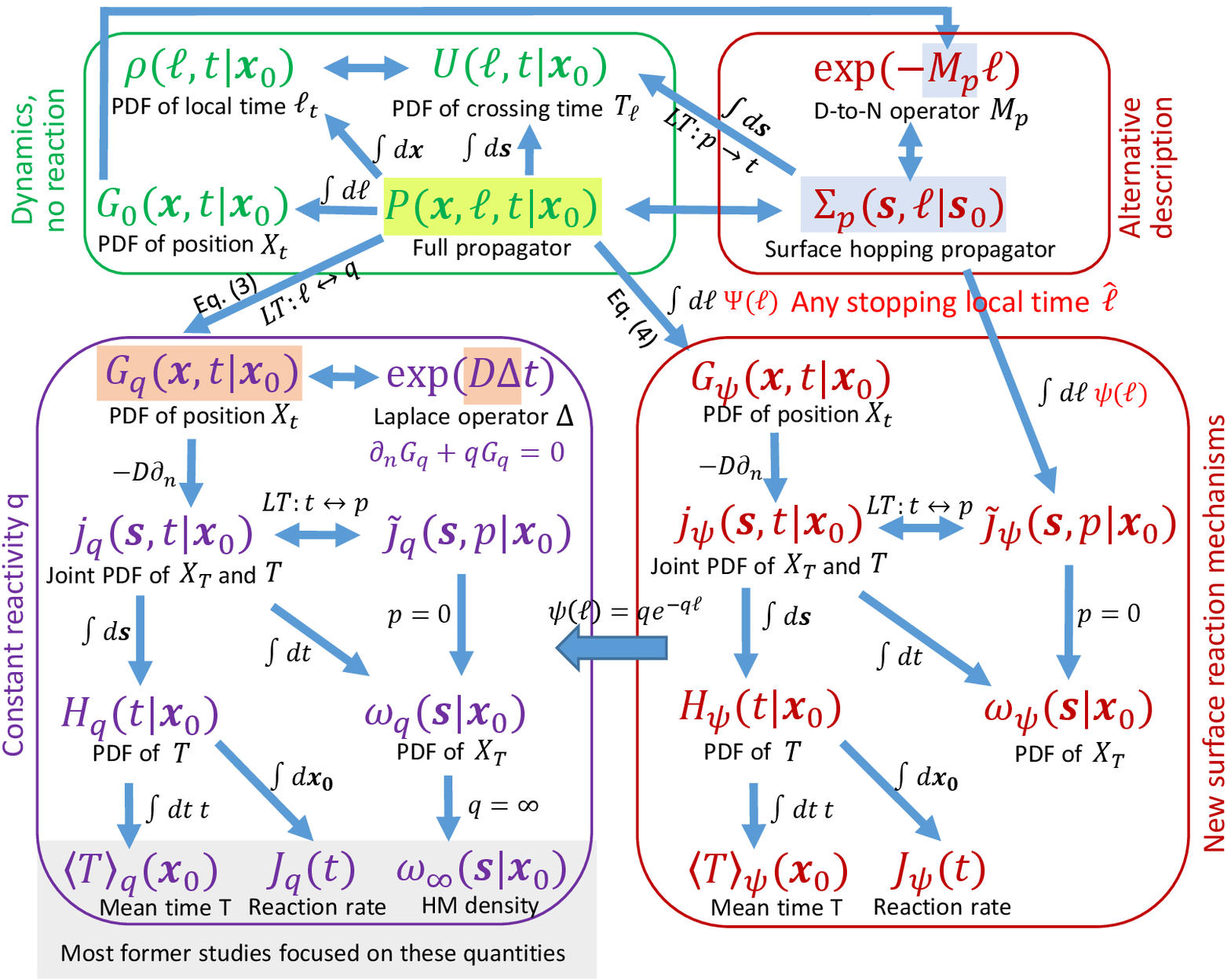} 
\end{center}
\caption{
Schematic view of interrelations between different functions
describing diffusion and reactions.  Top left frame includes the
functions that characterize diffusion alone, without any reaction
(reflecting boundary): the full propagator $P(\x,\ell,t|\x_0)$, the
marginal propagator $G_0(\x,t|\x_0)$ for zero reactivity, the
probability density function (PDF) $\rho(\ell,t|\x_0)$ of the boundary
local time $\ell_t$, and the PDF $U(\ell,t|\x_0)$ of the
first-crossing time $T_\ell$ of a level $\ell$ by $\ell_t$.  Top right
frame illustrates our alternative description based on the surface
hopping propagator $\Sigma_p(\s,\ell|\s_0)$ and the associated
Dirichlet-to-Neumann (D-to-N) operator $\M_p$, from which all other
quantities can be derived (see arrows).  Bottom left frame englobes
the functions within the conventional description of partially
reactive boundary by the Laplace operator $\Delta$, in which constant
reactivity $\kappa$ is implemented {\it implicitly} via Robin boundary
condition with parameter $q = \kappa/D$: the conventional propagator
$G_q(\x,t|\x_0)$, the probability flux density $j_q(\s,t|\x_0)$ and
its Laplace transform $\tilde{j}_q(\s,p|\x_0)$, the PDF $H_q(t|\x_0)$
of the reaction time $\T$ and the related survival probability, the
PDF $\omega_q(\s|\x_0)$ of the reaction location $\X_\T$ on the
boundary, the mean reaction time $\langle \T \rangle_q(\x_0)$, the
reaction rate $J_q(t)$, and the harmonic measure (HM) density
$\omega_\infty(\s|\x_0)$.  Bottom right frame presents all the above
quantities generalized to a variety of surface reaction mechanisms
determined {\it explicitly} by the stopping local time $\hat{\ell}$
with a given PDF $\psi(\ell)$.  Note that the Laplace transform (LT)
with respect to time $t$ allows one to incorporate bulk reactivity $p$
and thus to consider mortal walkers.  Detailed relations between
quantities are provided in Sec. \ref{sec:alternative} and
\ref{sec:generalized}.}
\label{fig:scheme}
\end{figure}

To facilitate reading, we summarize few basic notations: $\x$ and
$\x_0$ are bulk points (in $\bar{\Omega} = \Omega \cup \pa$); $\s$ and
$\s_0$ are boundary points (on $\pa$); $\hat{t}$ and $\hat{\ell}$ are
random variables; tilde denotes the Laplace transform with respect to
time $t$ (see also Fig. \ref{fig:scheme} for notations of many other
quantities).  We also outline a slight abuse in notations for boundary
points: in some functions, $\s$ and $\s_0$ are considered as points in
$\R^d$ lying on the boundary $\pa$; in other formulas, $\s$ and $\s_0$
are understood as boundary points lying on a lower-dimensional
manifold $\pa$.  For instance, for a half-plane $\R \times \R_+$, the
same notation will be used for a boundary point $s \in \R$ and for a
bulk point $(s,0) \in \R^2$ lying on the boundary.  The dimensionality
of such points clearly follows from the context; e.g., in
Eq. (\ref{eq:identity2}) below, $G_q(\s,t|\s_0)$ is the bulk
propagator with units m$^{-d}$, whereas $\Sigma_p(\s,\ell|\s_0)$ is
the surface hopping propagator with units m$^{1-d}$.

\subsection{Full propagator}

The starting point of the conventional description is the propagator
$G_q(\x,t|\x_0)$ that characterizes the likelihood of finding a
particle that started from a point $\x_0$ at time $0$ and not reacted
up to time $t$, in a bulk point $\x$ at time $t$.  For ordinary
diffusion in an Euclidean domain $\Omega\subset \R^d$ with a smooth
boundary $\pa$, the propagator satisfies the diffusion equation 
\begin{equation}  \label{eq:GqA}
\partial_t G_q(\x,t|\x_0) = D \Delta  G_q(\x,t|\x_0) \qquad (\x\in\Omega)
\end{equation}
for any fixed point $\x_0 \in \bar{\Omega}$, where $D$ is the
diffusion coefficient, and $\Delta$ is the Laplace operator
\cite{Risken}.  This equation describes how the uncertainty on the
location of the diffusing particle evolves with time from its initial
deterministic location at a fixed starting point $\x_0$:
$G_q(\x,t=0|\x_0) =
\delta(\x - \x_0)$, where $\delta(\x)$ is the Dirac distribution.
This equation is completed by the Robin boundary condition:
\begin{equation}  \label{eq:RobinA}
- \left.\bigl(\partial_n G_q(\x,t|\x_0)\bigr)\right|_{\x = \s} = q \, G_q(\s,t|\x_0)  \quad (\s \in \pa),
\end{equation}
where $\partial_n$ is the normal derivative directed outward the
domain $\Omega$, and $q = \kappa/D$.  For unbounded domains, a
regularity condition is also imposed at infinity: $G_q(\x,t|\x_0)\to
0$ as $|\x|\to \infty$.

In \cite{Grebenkov06,Grebenkov07a}, several constructions of partially
reflected Brownian motion, associated with the propagator
$G_q(\x,t|\x_0)$, were discussed (see also
\cite{Papanicolaou90,Bass08} for mathematical details and references).
In particular, one can start from reflected Brownian motion $\X_t$ in
$\Omega$ with {\it reflecting} boundary $\pa$, obtained as the
solution of the Skorokhod equation (\ref{eq:Skorokhod}), and stop it
at a random reaction time
\begin{equation} \label{eq:Tdef}
\T := \inf\{ t > 0 ~:~ \ell_t > \hat{\ell} \} ,
\end{equation}
when the boundary local time $\ell_t$ exceeds an independent random
variable $\hat{\ell}$ with the exponential distribution:
$\P\{\hat{\ell} > \ell\} = e^{- q \ell}$.  
In fact, as discussed in Sec. \ref{sec:local_time}, the boundary local
time $\ell_t$ is related via Eq. (\ref{eq:ell_def}) to the number
$\N_t^a$ of encounters with a thin boundary layer of width $a$.  At
each encounter with the partially reactive boundary, the particle
either reacts with the probability $\Pi = 1/(1 + D/(\kappa a)) \simeq
aq \ll 1$ (that follows from a discretization of Robin boundary
condition (\ref{eq:RobinA}) at small scale $a$, see
\cite{Filoche99,Grebenkov03}), or resumes its bulk diffusion with the
probability $1-\Pi$.  If all reaction attempts are independent from
each other, the number $\hat{n}$ of failed attempts until the
successful reaction follows the geometric law, $\P\{ \hat{n} > n\} =
(1 - \Pi)^n \approx \exp(-q na)$.  The rescaled number of failed
reaction attempts, $\hat{\ell} = a \hat{n}$, obeys thus the
exponential law: $\P\{ \hat{\ell} > \ell \} = e^{-q\ell}$, with $\ell
= an$.  As $\hat{\ell} = \ell_\T$ by construction, the reaction time
$\T$ is defined by Eq. (\ref{eq:Tdef}), see
\cite{Grebenkov06,Grebenkov07a,Grebenkov19c} for details.  After
this reminder of the probabilistic construction of the reaction time
$\T$, we are going to establish the fundamental representation
(\ref{eq:Pfull_Gq_SI}) for the conventional propagator, which to our
knowledge is new.

The conventional propagator is by definition the probability density
of the position $\X_t$ of the {\it survived} particle at time $t$,
i.e., of the particle that has not reacted up to time $t$:
\begin{equation}
G_q(\x,t|\x_0) d\x := \P_{\x_0}\{ \X_t \in (\x,\x+d\x), ~ t < \T\}.
\end{equation}
As the boundary local time $\ell_t$ is a nondecreasing process, the
condition $t < \T$ is equivalent to $\ell_t < \hat{\ell}$ according to
Eq. (\ref{eq:Tdef}).  In other words, one has
\begin{align*}
G_q(\x,t&|\x_0)  d\x = \P_{\x_0}\{ \X_t \in (\x,\x+d\x), ~ \ell_t < \hat{\ell}\} \\
& = \int\limits_0^\infty d\ell \, \underbrace{q \, e^{-q\ell}}_{\textrm{pdf of}~\hat{\ell}} \, \P_{\x_0}\{ \X_t \in (\x,\x+d\x), ~ \ell_t < \ell\} \\
& = \int\limits_0^\infty d\ell \, q \, e^{-q\ell} \, \int\limits_0^\ell d\ell' \, P(\x,\ell',t|\x_0) d\x \\
& = d\x \int\limits_0^\infty d\ell \, e^{-q\ell}\, P(\x,\ell,t|\x_0) ,
\end{align*}
i.e.,
\begin{equation}  \label{eq:Pfull_Gq_SI}
G_q(\x,t|\x_0) = \int\limits_0^\infty d\ell\, e^{-q\ell} \, P(\x,\ell,t|\x_0).
\end{equation}
Here we introduced the full propagator $P(\x,\ell,t|\x_0)$ of
reflected Brownian motion (without surface reaction), i.e., the joint
probability density of the position $\X_t$ and the boundary local time
$\ell_t$.  One sees that the exponential law of the stopping local
time $\hat{\ell}$ allowed us to represent the conventional propagator
as the Laplace transform of the full propagator with respect to the
local time $\ell$.

To proceed, we recall a useful representation of the conventional
propagator from \cite{Grebenkov19}:
\begin{align} \nonumber
& G_q(\x,t|\x_0) = G_\infty(\x,t|\x_0) +
\int\limits_{\pa} d\s_1 \int\limits_{\pa} d\s_2 \int\limits_0^t dt_1 \int\limits_{t_1}^t dt_2 \,  \\   \label{eq:Gq_identity}
& \times  j_{\infty}(\s_2,t-t_2|\x) \, G_q(\s_2,t_2-t_1|\s_1) \, j_{\infty}(\s_1,t_1|\x_0),
\end{align}
where%
\footnote{
We outline a confusing difference in notations between
\cite{Grebenkov19} and the present paper: in \cite{Grebenkov19}, the
subscript $0$ referred to the perfectly reactive boundary that we
denote here by the subscript $\infty$.}
%
\begin{equation}
j_{\infty}(\s,t|\x_0) := -\left. \bigl(D\partial_n G_\infty(\x,t|\x_0)\bigr)\right|_{\x=\s} 
\end{equation}
is the probability flux density on a perfectly reactive boundary ($q =
\infty$), i.e., with $G_\infty(\x,t|\x_0)$ satisfying Dirichlet boundary
condition: $G_\infty(\x,t|\x_0)|_{\x\in \pa} = 0$.  The representation
(\ref{eq:Gq_identity}) has a straightforward probabilistic
interpretation.  The first term describes the contribution of direct
trajectories from $\x_0$ to $\x$, which do not hit the boundary over
time from $0$ to $t$.  In turn, the second term provides the
contribution for indirect trajectories: the particle first arrives at
a boundary point $\s_1$ at time $t_1$ (factor
$j_{\infty}(\s_1,t_1|\x_0)$), then executes partially reflected
Brownian motion until time $t_2$ (factor $G_q(\s_2,t_2-t_1|\s_1)$), at
which the particle leaves the boundary from a point $\s_2$ and goes
directly to the bulk point $\x$ (factor $j_{\infty}(\s_2,t-t_2|\x)$).
Expectedly, one integrates over intermediate boundary points $\s_1$
and $\s_2$ and times $t_1$ and $t_2$.

The inverse Laplace transform of Eq. (\ref{eq:Gq_identity}) with
respect to $q$ yields, due to Eq. (\ref{eq:Pfull_Gq_SI}), a similar
representation for the full propagator:
\begin{align} \nonumber
& P(\x,\ell,t|\x_0) = G_\infty(\x,t|\x_0) \delta(\ell) + 
\int\limits_{\pa} \hspace*{-1mm} d\s_1 \hspace*{-1mm} \int\limits_{\pa} \hspace*{-1mm} d\s_2 \hspace*{-1mm}
\int\limits_0^t dt_1 \int\limits_{t_1}^t dt_2 \,  \\   \label{eq:PP_identity}
& \times j_{\infty}(\s_2,t-t_2|\x) \, P(\s_2,\ell,t_2-t_1|\s_1) \, j_{\infty}(\s_1,t_1|\x_0).
\end{align}
The probabilistic interpretation is again straightforward.  For a
direct trajectory from $\x_0$ to $\x$, which does not hit the
boundary, the boundary local time remains unchanged (the first term).
In turn, the increase of the boundary local time from $0$ to $\ell$
can only be achieved through indirect trajectories that hit the
boundary (the second term).  In this way, the full propagator for
arbitrary points $\x$ and $\x_0$ is expressed in terms of the full
propagator for boundary points $\s_1$ and $\s_2$. 

The Laplace transform of Eq. (\ref{eq:PP_identity}) with respect to
$t$ reduces time convolutions:
\begin{align}  \label{eq:Pfull3}
& \tilde{P}(\x,\ell,p|\x_0) = \tilde{G}_\infty(\x,p|\x_0) \delta(\ell) \\  \nonumber
& + \int\limits_{\pa} d\s_1 \int\limits_{\pa} d\s_2 \,
\tilde{j}_{\infty}(\s_2,p|\x) \, \tilde{P}(\s_2,\ell,p|\s_1) \, \tilde{j}_{\infty}(\s_1,p|\x_0),
\end{align}
where tilde denotes Laplace-transformed quantities (with respect to
$t$), e.g., 
\begin{equation}
\tilde{P}(\x,\ell,p|\x_0) := \int\limits_0^\infty dt \, e^{-pt} \, P(\x,\ell,t|\x_0).  
\end{equation}
In particular, for a boundary point $\x = \s \in \pa$, the first term
vanishes, whereas $\tilde{j}_{\infty}(\s_2,p|\s) = \delta(\s_2 -\s)$,
so that
\begin{align}  \label{eq:Pfull4}
& \tilde{P}(\s,\ell,p|\x_0) = \int\limits_{\pa} d\s_1 \, \tilde{P}(\s,\ell,p|\s_1)
\, \tilde{j}_{\infty}(\s_1,p|\x_0)  .
\end{align}

\subsection{Surface hopping propagator}

To express $\tilde{P}(\s_2,\ell,p|\s_1)$, we now introduce the surface
hopping propagator $\Sigma_p(\s,\ell|\s_0)$.  For this purpose, we
compute the marginal probability density of the reaction position
$\X_\T$ by two alternative ways.

On one hand, we define the surface hopping propagator
$\Sigma_p(\s,\ell|\s_0)$ as the probability density of finding a
particle, started from a boundary point $\s_0$ and survived bulk
reaction with a rate $p$, in a boundary point $\s$ at the local time
$\ell$.  Since the surface reaction occurs at a random, exponentially
distributed stopping local time $\hat{\ell}$, which is independent of
the particle's motion, the probability density of the reaction
position $\X_\T$ can be obtained by averaging
$\Sigma_p(\s,\hat{\ell}|\s_0)$ with the exponential density
$qe^{-q\ell}$ for $\hat{\ell}$:
\begin{equation} \label{eq:auxil77}
\int\limits_0^\infty d\ell \, q e^{-q\ell} \, \Sigma_p(\s,\ell|\s_0).
\end{equation}

On the other hand, the conventional propagator determines the
probability flux density at a boundary point $\s\in\pa$:
\begin{equation}  \label{eq:jq_def}
j_q(\s,t|\x_0) := - D \left. \bigl(\partial_n G_q(\x,t|\x_0)\bigr)\right|_{\x=\s} \,,
\end{equation}
i.e., the joint probability density of the stopping position $\X_\T$
on the boundary and of the stopping time $\T$:
\begin{equation}
j_q(\s,t|\x_0) \, d\s \, dt = \P_{\x_0}\{ \X_\T\in(\s,\s+d\s), ~ \T \in (t,t+dt)\} \,.
\end{equation}
Note that Robin boundary condition (\ref{eq:RobinA}) yields an
equivalent relation
\begin{equation}  \label{eq:jq_def2}
j_q(\s,t|\x_0) = q D G_q(\s,t|\x_0) \,.
\end{equation}
For a particle with a finite random lifetime $\hat{t}$ due to bulk
reaction, one needs to include the additional condition that the
particle has reacted on the surface during its lifetime (i.e., before
its death): $\T < \hat{t}$.  Since $\hat{t}$ is independent from both
$\X_t$ and $\T$, one has
\begin{align*}
& \P_{\x_0}\{ \X_\T\in(\s,\s+d\s),~ \T \in (t,t+dt), ~ \T < \hat{t} \}  \\
& = \P\{\hat{t} > t \} \, j_q(\s,t|\x_0) \, dt \, d\s  ,
\end{align*}
where $\P\{ \hat{t} > t\} = e^{-pt}$ for the considered
Poissonian-type bulk reaction with the rate $p$.  The integral of this
probability density over $t$ yields the marginal probability density
of the reaction position $\X_{\T}$ (for the particle that survived
bulk reaction), which turns out to coincide with the Laplace transform
of $j_q(\s,t|\x_0)$:
\begin{equation}
\tilde{j}_q(\s,p|\x_0) := \int\limits_0^\infty dt \, e^{-pt} \, j_q(\s,t|\x_0).
\end{equation}
When the starting point $\x_0$ lies on the boundary, $\x_0 = \s_0\in
\pa$, $\tilde{j}_q(\s,p|\s_0)$ is identical, by construction, to the
probability density of $\X_\T$ given by Eq. (\ref{eq:auxil77}):
\begin{equation}  \label{eq:identity}
\tilde{j}_q(\s,p|\s_0) = \int\limits_0^\infty d\ell \, q \, e^{-q\ell} \, \Sigma_p(\s,\ell|\s_0).
\end{equation}
Expressing the left-hand side with the aid of Eq. (\ref{eq:jq_def2}),
we can re-write this relation more explicitly in terms of two Laplace
transforms:
\begin{equation}  \label{eq:identity2}
D \int\limits_0^\infty dt \, e^{-pt} \, G_q(\s,t|\s_0) = \int\limits_0^\infty d\ell \, e^{-q\ell} \, \Sigma_p(\s,\ell|\s_0) .
\end{equation}
This fundamental identity relates the conventional propagator
$G_q(\s,t|\s_0)$ to the surface hopping propagator
$\Sigma_p(\s,\ell|\s_0)$.  This relation highlights the duality
between bulk and surface reaction mechanisms which involve time $t$
and local time $\ell$ via respective Laplace transforms.  Even so
Eq. (\ref{eq:identity2}) determines $G_q(\s,t|\s_0)$ only at boundary
points $\s$ and $\s_0$, the propagator $G_q(\x,t|\x_0)$ can then be
recovered from Eq. (\ref{eq:Gq_identity}) for all bulk points $\x$ and
$\x_0$.

Comparing the identity (\ref{eq:identity2}) to
Eq. (\ref{eq:Pfull_Gq_SI}), evaluated at $\x=\s$ and $\x_0=\s_0$, we
also conclude that
\begin{equation}  \label{eq:Sigma_P2}
\Sigma_p(\s,\ell|\s_0) = D \int\limits_0^\infty dt \, e^{-pt} \, P(\s,\ell,t|\s_0)  .
\end{equation}
Bearing in mind this relation, one finally realizes that the identity
(\ref{eq:identity2}) reflects a very simple statement: if there are
both bulk and surface reaction mechanisms, the order of implementation
of these mechanisms does not matter.  In fact, the conventional
approach consists in incorporating first the surface reactivity via
the Robin boundary condition to get $G_q(\s,t|\s_0)$ and then
averaging it with the survival probability $\P\{\hat{t} > t\} =
e^{-pt}$ to include bulk reaction (the left-hand side of
Eq. (\ref{eq:identity2})).  In turn, the alternative description
proposed here consists in implementing first the bulk reactivity via
Eq. (\ref{eq:Sigma_P2}) and then averaging the surface hopping
propagator $\Sigma_p(\s,\ell|\s_0)$ with the probability density $q
e^{-q\ell}$ of the stopping local time $\hat{\ell}$ to include surface
reaction (the right-hand side of Eq. (\ref{eq:identity2})).  Note also
that the inverse Laplace transform of Eq. (\ref{eq:Sigma_P2}) with
respect to $p$ reads
\begin{equation}  \label{eq:Pt_Sigmap}
P(\s,\ell,t|\s_0) = \frac{1}{D} \L^{-1}_t \bigl\{\Sigma_p(\s,\ell|\s_0) \bigr\}. 
\end{equation}

Substitution of Eq. (\ref{eq:Sigma_P2}) into Eq. (\ref{eq:Pfull3})
yields
\begin{align}  \label{eq:Pfull30}
& \tilde{P}(\x,\ell,p|\x_0) = \tilde{G}_\infty(\x,p|\x_0) \delta(\ell) \\  \nonumber
& + \int\limits_{\pa} d\s_1 \int\limits_{\pa} d\s_2 \,
\tilde{j}_{\infty}(\s_2,p|\x) \, \frac{\Sigma_p(\s_2,\ell|\s_1)}{D} \, \tilde{j}_{\infty}(\s_1,p|\x_0),
\end{align}
i.e., the surface hopping propagator, together with
$\tilde{G}_\infty(\x,p|\x_0)$, determines the full propagator
$P(\x,\ell,t|\x_0)$ in the Laplace domain.  This relation is the basis
for computing the full propagator.

\subsection{Dirichlet-to-Neumann operator}

The last step consists in deriving the spectral decomposition of the
surface hopping propagator on the basis of the Dirichlet-to-Neumann
operator $\M_p$.  For a given function $f$ on the boundary $\pa$, the
operator $\M_p$ associates another function $g = \M_p f = (\partial_n
w)|_{\pa}$, where $w$ is the solution of the Dirichlet boundary value
problem for the modified Helmholtz equation:
\begin{equation}  \label{eq:u_def}
(p - D\Delta) w = 0 \quad (\x\in\Omega), \qquad w|_{\pa} = f 
\end{equation}
(with the regularity condition $w(\x)\to 0$ as $|\x|\to \infty$ if
$\Omega$ is not bounded).  The Dirichlet-to-Neumann operator $\M_p$ is
a self-adjoint pseudo-differential operator (see
\cite{Arendt14,Daners14,Arendt15,Hassell17,Girouard17}).
On one hand, the action of the Dirichlet-to-Neumann operator on a
function $f(\s)$ on the boundary can be expressed by solving the
boundary value problem (\ref{eq:u_def}) in a standard way with the
help of the Laplace-transformed propagator
$\tilde{G}_{\infty}(\x,p|\x_0)$ with Dirichlet boundary condition ($q
= \infty$):
\begin{align}  \label{eq:Mp_def1}
& [\M_p f](\s_0) \\  \nonumber
&= \biggl(\partial_{n_0} \int\limits_{\pa} d\s \, 
\underbrace{\left. \bigl(-D \partial_n \tilde{G}_{\infty}(\x,p|\x_0)\bigr)\right|_{\x=\s}}_{ = \tilde{j}_\infty(\s,p|\x_0)} \, f(\s)\biggr)_{\x_0 = \s_0} .
\end{align}
On the other hand, the inverse of the Dirichlet-to-Neumann operator
$\M_p$ with $p > 0$ can be expressed in terms of the
Laplace-transformed propagator $\tilde{G}_{0}(\x,p|\x_0)$ with Neumann
boundary condition ($q = 0$) \cite{Grebenkov19}:
\begin{equation}  \label{eq:Mp_inv}
D \tilde{G}_{0}(\s,p|\s_0) = \M_p^{-1} \delta(\s - \s_0)  \quad (\s,\s_0\in \pa).
\end{equation}
More generally, $\tilde{G}_{q}(\s,p|\s_0)$ was shown to be related to
the resolvent of the Dirichlet-to-Neumann operator \cite{Grebenkov19}:
\begin{equation} \label{eq:Mresolvent}
D\tilde{G}_q(\s,p|\s_0) = (qI + \M_p)^{-1} \delta(\s-\s_0),
\end{equation}
where $I$ is the identity operator.  Substituting this expression into
Eq. (\ref{eq:identity2}) and performing the inverse Laplace transform
with respect to $q$, we show that 
\begin{equation}  \label{eq:Sigma_semi}
\Sigma_p(\s,\ell|\s_0) = \exp(-\M_p \ell) \delta(\s-\s_0),
\end{equation}
i.e., the surface hopping propagator is the kernel of the semigroup
$\exp(-\M_p \ell)$ associated with the Dirichlet-to-Neumann operator
$\M_p$.  Taking the derivative with respect to $\ell$, one immediately
gets 
\begin{equation}    \label{eq:Sigma_eq}
\partial_\ell \Sigma_p(\s,\ell|\s_0) = - \M_p \, \Sigma_p(\s,\ell|\s_0),
\end{equation}
subject to the initial condition with Dirac distribution: 
\begin{equation}   \label{eq:Sigma_ini}
\Sigma_p(\s,\ell=0|\s_0) = \delta(\s-\s_0).  
\end{equation}
This is a sort of diffusion equation characterizing surface
exploration by bulk-mediated diffusion hops on the boundary, whose
``number'' is quantified by the boundary local time $\ell$, in analogy
with bulk jumps of a random walk quantified by the physical time $t$.

When the boundary $\pa$ is bounded, the Dirichlet-to-Neumann operator
$\M_p$ has a discrete spectrum, i.e., a countable set of eigenvalues
$\mu_n^{(p)}$ and eigenfunctions $v_n^{(p)}(\s)$ satisfying
\begin{equation}  \label{eq:M_eigen}
\M_p \, v_n^{(p)}(\s) = \mu_n^{(p)} \, v_n^{(p)}(\s) .
\end{equation}
The eigenvalues are nonnegative, whereas the eigenfunctions form an
orthonormal complete basis in the space $L_2(\pa)$ of
square-integrable functions on the boundary $\pa$.  We emphasize that
this property holds irrespectively of whether the domain $\Omega$ is
bounded or not.
In the bounded case, Eqs. (\ref{eq:Mresolvent}, \ref{eq:Sigma_semi})
imply the spectral decompositions:
\begin{equation}
D \tilde{G}_q(\s,p|\s_0) = \sum\limits_{n} [v_n^{(p)}(\s_0)]^* \, v_n^{(p)}(\s) \, \frac{1}{q + \mu_n^{(p)}} 
\end{equation}
and
\begin{equation}  \label{eq:Sigmap_spectral}
\Sigma_p(\s,\ell|\s_0) = \sum\limits_{n} [v_n^{(p)}(\s_0)]^* \, v_n^{(p)}(\s) \, \exp(- \mu_n^{(p)} \ell) \,,
\end{equation}
where asterisk denotes the complex conjugate.  The completeness of the
eigenfunctions $v_n^{(p)}(\s)$ ensures the correct initial condition
(\ref{eq:Sigma_ini}).  Substituting Eq. (\ref{eq:Sigmap_spectral})
into Eq. (\ref{eq:Pfull30}), we derive the spectral expansion of the
Laplace-transformed full propagator:
\begin{align}  \label{eq:Sigmap_spectral2}
\tilde{P}(\x,\ell,p|\x_0) & = \tilde{G}_\infty(\x,p|\x_0) \, \delta(\ell)  \\  \nonumber
& + \frac{1}{D} \sum\limits_{n} [V_n^{(p)}(\x_0)]^* \, V_n^{(p)}(\x) \, \exp(- \mu_n^{(p)} \ell) \,,
\end{align}
where
\begin{equation}  \label{eq:Vnp}
V_n^{(p)}(\x_0) := \int\limits_{\pa} d\s_0 \, v_n^{(p)}(\s_0) \, \tilde{j}_\infty(\s_0,p|\x_0)
\end{equation}
is the projection of $\tilde{j}_\infty(\s_0,p|\x_0)$ onto the
eigenfunction $v_n^{(p)}$ of the Dirichlet-to-Neumann operator.  By
construction, each function $V_n^{(p)}(\x)$ satisfies the modified
Helmholtz equation (\ref{eq:u_def}) with Dirichlet boundary condition,
$(V_n^{(p)})_{|\pa} = v_n^{(p)}$, i.e., they naturally appear in the
definition of the Dirichlet-to-Neumann operator $\M_p$; in particular,
$(\partial_n V_n^{(p)})_{|\pa} = \mu_n^{(p)} \, v_n^{(p)}$.

The profound relation (\ref{eq:identity2}) between the conventional
propagator and the surface hopping propagator allows one to derive new
spectral decompositions for many quantities characterizing
diffusion-influenced reactions.  For instance,
Eq. (\ref{eq:identity}), written more generally for a bulk starting
point $\x_0$ (instead of $\s_0$), yields
\begin{equation}  \label{eq:tildejq}
\tilde{j}_q(\s,p|\x_0) = \sum\limits_{n} [V_n^{(p)}(\x_0)]^* \, v_n^{(p)}(\s) \, \frac{q}{q + \mu_n^{(p)}} \,,
\end{equation}
which was earlier reported in \cite{Grebenkov19}.  Other spectral
expansions will be provided in Sec. \ref{sec:generalized}.

\subsection{Boundary value problem for the full propagator}

The full propagator $P(\x,\ell,t|\x_0)$ is determined in the Laplace
domain by Eq. (\ref{eq:Pfull30}), which depends on the surface hopping
propagator and thus can be computed from the spectral properties of
the Dirichlet-to-Neumann operator.  Nevertheless, it is instructive to
formulate the boundary value problem for the full propagator.

As the first term of the Skorokhod equation (\ref{eq:Skorokhod})
affects reflected Brownian motion only on the boundary, the full
propagator is expected to obey the ordinary diffusion equation in the
bulk
\begin{equation}  \label{eq:P_eq}
\partial_t P(\x,\ell,t|\x_0) = D \Delta P(\x,\ell,t|\x_0),
\end{equation}
subject to the initial condition 
\begin{equation}  \label{eq:P_initial}
P(\x,\ell,t=0|\x_0) = \delta(\x-\x_0) \delta(\ell).
\end{equation}
This property can be checked by applying the operator $(p - D\Delta)$
to Eq. (\ref{eq:Pfull3}) that yields
\begin{equation}  \label{eq:auxil57}
(p-D\Delta) \tilde{P}(\x,\ell,p|\x_0) = \delta(\x-\x_0) \delta(\ell) .
\end{equation}
In fact, the second term of Eq. (\ref{eq:Pfull3}) vanishes since
$(p-D\Delta) \tilde{j}_\infty(\s_2,p|\x) = 0$ for any bulk point
$\x\in\Omega$, and we used that $(p-D\Delta)
\tilde{G}_\infty(\x,p|\x_0) = \delta(\x-\x_0)$.  The inverse Laplace
transform of Eq. (\ref{eq:auxil57}) implies Eqs. (\ref{eq:P_eq},
\ref{eq:P_initial}).

In turn, the boundary local time increases at each encounter with the
boundary, so that its effect should be taken into account via an
appropriate boundary condition.  As changes of the full propagator
with respect to $\ell$ are driven by arrivals of the particle onto the
boundary, the derivative of $P(\x,\ell,t|\x_0)$ with respect to $\ell$
is expected to be related to the normal derivative of the full
propagator on the boundary.  To establish this relation, we first
evaluate the normal derivative of the Laplace-transformed full
propagator in Eq. (\ref{eq:Pfull30}):
\begin{align} 
& \left. -D\bigl(\partial_n \tilde{P}(\x,\ell,p|\x_0)\bigr)\right|_{\x = \s} = \tilde{j}_\infty(\s,p|\x_0) \, \delta(\ell) \\   \nonumber
&  -  \int\limits_{\pa} d\s_1 \, \tilde{j}_\infty(\s_1,p|\x_0) \\  \nonumber
& \qquad \times \biggl. \biggl(  \partial_n  \int\limits_{\pa} d\s_2 \, \tilde{j}_\infty(\s_2,p|\x) \,
  \Sigma_p(\s_2,\ell|\s_1)   \biggr) \biggr|_{\x=\s} .
\end{align}
According to Eq. (\ref{eq:Mp_def1}), the factor in the last line can
be interpreted as the action of the Dirichlet-to-Neumann operator
$\M_p$ on $\Sigma_p(\cdot,\ell|\s_1)$; in turn, Eqs. (\ref{eq:Pfull4},
\ref{eq:Sigma_eq}) imply
\begin{align}  \nonumber
\left. -D\bigl(\partial_n \tilde{P}(\x,\ell,p|\x_0)\bigr)\right|_{\x = \s} & = \tilde{j}_\infty(\s,p|\x_0) \, \delta(\ell) \\  \label{eq:relation_aux12}
& + D\partial_\ell \tilde{P}(\s,\ell,p|\x_0)   .
\end{align}
The inverse Laplace transform of this relation with respect to $p$
yields the desired boundary condition for the full propagator:
\begin{align}  \nonumber
\left. -D\bigl(\partial_n P(\x,\ell,t|\x_0)\bigr)\right|_{\x = \s} & = j_\infty(\s,t|\x_0) \, \delta(\ell) \\  \label{eq:P_BC}
& + D \partial_\ell P(\s,\ell,t|\x_0)   .
\end{align}
In addition, one has
\begin{equation}  \label{eq:jinfty_P}
P(\s,\ell=0,t|\x_0) = \frac{1}{D} \, j_\infty(\s,t|\x_0) \quad (\s\in\pa),
\end{equation}
which follows from the inverse Laplace transform of
Eq. (\ref{eq:Sigmap_spectral2}) evaluated at $\x = \s \in \pa$ and
$\ell = 0$ and using Eq. (\ref{eq:Sigma_ini}).

To complete the formulation, one also needs to impose the regularity
condition
\begin{equation} \label{eq:P_regul}
\lim\limits_{\ell\to\infty} P(\x,\ell,t|\x_0) = 0,
\end{equation}
ensuring that the boundary local time cannot be infinite within a
finite time $t$.  For unbounded domains, a similar regularity
condition on $\x$ is imposed:
\begin{equation}   \label{eq:P_regul2}
\lim\limits_{|\x|\to\infty} P(\x,\ell,t|\x_0) = 0.
\end{equation}
The boundary value problem (\ref{eq:P_eq}, \ref{eq:P_initial},
\ref{eq:P_BC}, \ref{eq:jinfty_P}, \ref{eq:P_regul}, \ref{eq:P_regul2})
determines the full propagator $P(\x,\ell,t|\x_0)$.

Multiplying the boundary condition (\ref{eq:P_BC}) by $e^{-q\ell}$,
integrating over $\ell$ from $0$ to $\infty$, and performing
integration by parts in the last term, one retrieves Robin boundary
condition (\ref{eq:RobinA}) for the conventional propagator
$G_q(\x,t|\x_0)$, where we used Eq. (\ref{eq:jinfty_P}).

\subsection{Distribution of the boundary local time}
\label{sec:BLT}

By definition, the integral of the full propagator $P(\x,\ell,t|\x_0)$
over the arrival point $\x$ gives the marginal probability density of
the boundary local time $\ell_t$:
\begin{equation}  \label{eq:rho_def}
\rho(\ell,t|\x_0) := \int\limits_\Omega d\x \, P(\x,\ell,t|\x_0).
\end{equation}
The properties of the boundary local time were discussed in
Sec. \ref{sec:local_time}.

The probability density in Eq. (\ref{eq:rho_def}) can be expressed in
terms of the surface hopping propagator.  For this purpose, it is
convenient to work in the Laplace domain, in which the integral of
Eq. (\ref{eq:Pfull30}) over $\x$ yields
\begin{align}  \label{eq:auxil23}
& \tilde{\rho}(\ell,p|\x_0) = \tilde{S}_\infty(p|\x_0) \delta(\ell) \\  \nonumber
& + \int\limits_{\pa} d\s_1 \int\limits_{\pa} d\s_2 \,
\frac{D}{p} [\M_p 1](\s_2) \, \frac{\Sigma_p(\s_2,\ell|\s_1)}{D}  \, \tilde{j}_{\infty}(\s_1,p|\x_0),
\end{align}
where we used the identity 
\begin{equation}
\int\limits_\Omega d\x \, \tilde{j}_\infty(\s_2,p |\x) = \frac{D}{p} [\M_p 1](\s_2)
\end{equation}
derived in \cite{Grebenkov19}, and 
\begin{equation}
\tilde{S}_\infty(p|\x_0) := \int\limits_\Omega d\x \, \tilde{G}_\infty(\x,p|\x_0)
\end{equation}
is the Laplace-transformed survival probability for a perfectly
reactive boundary.  Writing the second integral in
Eq. (\ref{eq:auxil23}) in the form of a scalar product, one has
\begin{align*}
& \bigl(\Sigma_p(\cdot,\ell|\s_1) \cdot \M_p 1\bigr)_{L_2(\pa)} = \bigl(\M_p \Sigma_p(\cdot,\ell|\s_1) \cdot 1\bigr)_{L_2(\pa)} \\
& = \bigl( -\partial_\ell \Sigma_p(\cdot,\ell|\s_1) \cdot 1\bigr)_{L_2(\pa)} 
= -\partial_\ell \int\limits_{\pa} d\s_2 \,\Sigma_p(\s_2,\ell|\s_1) ,
\end{align*}
where we used Eq. (\ref{eq:Sigma_eq}).
The integral of $\Sigma_p(\s_2,\ell|\s_1)$ over $\s_2$ is the
probability that the particle is survived the bulk reaction up to the
local time $\ell$.  In other words, this is the probability that the
boundary local time $\ell_t$, stopped at $t = \hat{t}$, exceeds
$\ell$:
\begin{align}  \nonumber
\int\limits_{\pa} d\s_2 \,\Sigma_p(\s_2,\ell|\s_1) &= \P_{\s_1}\{\ell_{\hat{t}} > \ell\} \\
&= \int\limits_0^\infty dt \, p \, e^{-pt} \, \P_{\s_1}\{\ell_t > \ell\}.
\end{align}
We get thus
\begin{align*}
& \bigl(\Sigma_p(\cdot,\ell|\s_1) \cdot \M_p 1\bigr)_{L_2(\pa)} 
 = -\partial_\ell \int\limits_0^\infty dt \, p \, e^{-pt} \, \P_{\s_1}\{\ell_t > \ell\} \\
& = p \, \tilde{\rho}(\ell,p|\s_1),
\end{align*}
from which
\begin{equation}  \label{eq:tilde_rho_s0}
\tilde{\rho}(\ell,p|\s_0) = - \partial_\ell \frac{1}{p}  \int\limits_{\pa} d\s \, \Sigma_p(\s,\ell|\s_0).
\end{equation}
As a consequence, Eq. (\ref{eq:auxil23}) is reduced to
\begin{align}   \label{eq:tilde_rho_x0}
& \tilde{\rho}(\ell,p|\x_0) = \tilde{S}_\infty(p|\x_0) \delta(\ell) 
 + \int\limits_{\pa} d\s \, \tilde{\rho}(\ell,p|\s)  \, \tilde{j}_{\infty}(\s,p|\x_0),
\end{align}
which was reported in \cite{Grebenkov19c}.

Moreover, the Laplace transform of Eq. (\ref{eq:rho_def}) with respect
to $\ell$, together with Eq. (\ref{eq:Pfull_Gq_SI}), yields
\begin{equation}
\E_{\x_0}\{ e^{-q\ell_t} \} = \int\limits_0^\infty d\ell \, e^{-q\ell} \, \rho(\ell,t|\x_0) = S_q(t|\x_0),
\end{equation}
where 
\begin{equation}
S_q(t|\x_0) := \int\limits_\Omega d\x \, G_q(\x,t|\x_0)
\end{equation}
is the survival probability for partially reactive boundary.  This
relation allows one to interpret the survival probability
$S_q(t|\x_0)$ as the moment-generating function of the boundary local
time $\ell_t$.
Similarly, Eq. (\ref{eq:Pfull_Gq_SI}) provides an interpretation of
the conventional propagator $G_q(\x,t|\x_0)$ as the moment-generating
function of the boundary local time $\ell_t$ with the constraint on
the position $\X_t$:
\begin{equation}
G_q(\x,t|\x_0) = \E_{\x_0}\bigl\{ \exp(-q\ell_t) \, \delta(\X_t - \x) \bigr\} .
\end{equation}
To our knowledge, these representations were not reported
earlier.

It is also instructive to reconsider the approximation $\ell_t^a$ of
the boundary local time $\ell_t$ as the residence time in a thin
boundary layer $\pa_a$ of width $a$, rescaled by $D/a$, see
Eq. (\ref{eq:ell_ta}).  We introduce its moment-generating function
with the constraint on the position $\X_t$:
\begin{equation}
G_q^{a}(\x,t|\x_0) = \E_{\x_0}\bigl\{ \exp(-q\ell_t^{a}) \, \delta(\X_t - \x) \bigr\} .
\end{equation}
According to the Feynman-Kac formula \cite{Freidlin}, this function
satisfies the backward Fokker-Planck equation, which can be written in
the forward form due to the time reversal symmetry:
\begin{equation}
\partial_t G_q^{a}(\x,t|\x_0) = \biggl(D \Delta - \frac{qD}{a} \,\I_{\pa_a}(\x) \biggr) G_q^{a}(\x,t|\x_0) ,
\end{equation}
subject to the initial condition $G_q^{a}(\x,t|\x_0) = \delta(\x -
\x_0)$ and Neumann boundary condition $(\partial_n
G_q^{a}(\x,t|\x_0))|_{\x\in\pa} = 0$.  This equation can be
interpreted as diffusion in $\Omega$ with reflecting boundary subject
to {\it bulk} reaction in the thin boundary layer $\pa_a$ with the
rate $qD/a = \kappa/a$.  In the limit $a\to 0$, $\ell_t^{a}$
approaches $\ell_t$ due to Eq. (\ref{eq:ell_res}) so that
$G_q^{a}(\x,t|\x_0)$ is expected to approach the conventional
propagator $G_q(\x,t|\x_0)$.  The equivalence of two descriptions was
earlier discussed in \cite{Wilemski73}.  It also resembles the basic
idea of the diffuse layer approximation, in which a sharp boundary is
replaced by a diffuse layer in order to represent boundary conditions
by appropriate bulk terms \cite{Li09,Yu12}.

\subsection{Distribution of the first-crossing time}
\label{sec:SI_U}

As surface reaction occurs at a random time $\T$ when the boundary
local time $\ell_t$ exceeds the random stopping local time
$\hat{\ell}$, it is natural to look at the distribution of the
first-crossing time $T_\ell$ of a fixed threshold $\ell$.

Let $U(\ell,t|\x_0)$ denote the probability density of the first
moment $T_\ell$ when the boundary local time $\ell_t$ exceeds $\ell$:
\begin{equation}
T_\ell := \inf\{ t>0 ~:~ \ell_t > \ell \}.
\end{equation}
Since the boundary local time is a nondecreasing process, one has
\begin{equation}
\P_{\x_0} \{ \ell_t > \ell \} = \P_{\x_0} \{ t > T_\ell\} ,
\end{equation}
from which the probability density of $T_\ell$ follows:
\begin{equation}
U(\ell,t|\x_0) = \partial_t \P_{\x_0} \{ t > T_\ell \} = \partial_t \int\limits_{\ell}^\infty d\ell' \, \rho(\ell',t|\x_0),
\end{equation}
where $\rho(\ell,t|\x_0)$ is the probability density of the boundary
local time $\ell_t$.  Applying the Laplace transform with respect to
$t$ and using Eqs. (\ref{eq:tilde_rho_s0}, \ref{eq:tilde_rho_x0}) for
$\tilde{\rho}(\ell,p|\x_0)$, we get
\begin{equation}  \label{eq:Utilde_Sigma}
\tilde{U}(\ell,p|\x_0) = p \int\limits_{\ell}^\infty d\ell' \, \tilde{\rho}(\ell',p|\x_0) 
= \int\limits_{\pa} d\s \, D\tilde{P}(\s,\ell,p|\x_0) .
\end{equation}
The spectral expansion (\ref{eq:Sigmap_spectral2}) yields
\begin{equation} \label{eq:U2}
\tilde{U}(\ell,p|\x_0) = \sum\limits_n e^{-\mu_n^{(p)} \ell} \, [V_n^{(p)}(\x_0)]^* \int\limits_{\pa} d\s \, v_n^{(p)}(\s)  .
\end{equation}

The inverse Laplace transform of Eq. (\ref{eq:Utilde_Sigma}) with
respect to $p$ also gives
\begin{equation}
U(\ell,t|\x_0) = D \int\limits_{\pa} d\s  \, P(\s,\ell,t|\x_0) .
\end{equation}
While the integral of the full propagator $P(\x,\ell,t|\x_0)$ over
bulk points $\x\in\Omega$ in Eq. (\ref{eq:rho_def}) yields the
marginal probability density $\rho(\ell,t|\x_0)$ of the boundary local
time, its integral over boundary points $\s \in \pa$, multiplied by
$D$, determines the probability density $U(\ell,t|\x_0)$ of the
first-crossing time $T_\ell$.

The density $U(\ell,t|\x_0)$ characterizes the dynamics of a diffusing
particle near a fully reflecting boundary (without any reaction) and
will thus play the central role in the analysis of the reaction time
(see Sec. \ref{sec:generalized}).  For instance, as the first crossing
of the level $\ell =0$ corresponds to the first arrival of the
particle onto the boundary, $U(0,t|\x_0)$ is the probability density
of the first-passage time to the boundary.  Most former studies of
first-passage properties were limited to this quantity, which
characterizes a perfectly reactive boundary.

\section{Encounter-dependent reactivity}  
\label{sec:passivation}

In the conventional description, at each encounter with a partially
reactive surface, the particle either reacted with a constant
probability $\Pi \approx a \kappa/D \ll 1$, or resumed its bulk
diffusion with probability $1-\Pi$.  We extend this description to
account for variable, encounter-dependent reactivity that can model
progressive passivation/activation or aging of the reactive surface
after each reaction attempt.  We consider now that, at $n$-th
encounter with the boundary, the reaction probability is $\Pi_n
\approx a \kappa_n/D$, with a given infinite sequence of reactivities
$\kappa_n$.  If reaction attempts are independent, then the
probability of reaction exactly after $n$ failed attempts is
\begin{equation}
\P\{ \hat{n} = n \} = (1-\Pi_0)(1-\Pi_1) \ldots (1-\Pi_{n-1}) \Pi_n ,
\end{equation}
(with $n=0,1,2,\ldots$), whereas the probability of at least $n$
failed attempts is
\begin{equation}
\P\{ \hat{n} \geq n \} = (1-\Pi_0)(1-\Pi_1) \ldots (1-\Pi_{n-1}) .
\end{equation}
Setting $\hat{\ell} = a\hat{n}$ and $\ell = an$, we get in the limit
$a\to 0$:
\begin{equation}  \label{eq:P_K}
\Psi(\ell) := \P\{ \hat{\ell} > \ell \} = \exp(- K(\ell)),
\end{equation}
where
\begin{equation}  \label{eq:Kell}
K(\ell) := \frac{1}{D} \int\limits_0^\ell d\ell' \, \kappa(\ell') ,
\end{equation}
with $\kappa(\ell)$ being the reactivity at the local time $\ell$,
i.e., an appropriate limit of $\kappa_n$.  The reactivity
$\kappa(\ell)$ should be a non-negative function which is integrable
in the vicinity of $0$ to ensure $\Psi(0) = 1$ (i.e., $\kappa(\ell)$
cannot diverge faster than $\ell^{-\nu}$ as $\ell\to 0$, with $\nu <
1$).  When $\kappa(\ell)$ is a constant $\kappa_0$, one gets $K(\ell)
= \kappa_0 \ell/D = q\ell$ and thus retrieves the exponential
distribution of the stopping local time $\hat{\ell}$.

Taking the derivative of Eq. (\ref{eq:P_K}), we get the probability
density of the stopping local time $\hat{\ell}$:
\begin{equation}  \label{eq:psi_kappa}
\psi(\ell) = \frac{\kappa(\ell)}{D} \exp\biggl(-\frac{1}{D}\int\limits_0^\ell d\ell' \, \kappa(\ell')\biggr).
\end{equation}
Conversely, inverting Eq. (\ref{eq:psi_kappa}), we express the
reactivity $\kappa(\ell)$ as
\begin{equation}  \label{eq:kappa_psi}
\kappa(\ell) = D \frac{\psi(\ell)}{\int\nolimits_\ell^\infty d\ell' \, \psi(\ell')} \,.
\end{equation}
As a consequence, surface reaction mechanism can be selected by
setting either the reactivity $\kappa(\ell)$, or the probability
density $\psi(\ell)$ of the stopping local time $\hat{\ell}$.

As discussed in the main text, the asymptotic behavior of
$\kappa(\ell)$ at large $\ell$ distinguishes three scenarios:

(i) If $\kappa(\ell)$ decreases slower than $1/\ell$ (or increases),
then $K(\ell)$ diverges as a power law or faster, $\psi(\ell)$ in
Eq. (\ref{eq:psi_kappa}) exhibits a fast (typically
stretched-exponential) decay in the limit $\ell \to\infty$, and thus
all positive moments of $\ell$ are finite, as in the conventional
setting of a constant $\kappa$.

(ii) In the critical regime when $\kappa(\ell)$ decays as $\nu D/\ell$
(with some dimensionless constant $\nu > 0$), Eq. (\ref{eq:psi_kappa})
implies that $\psi(\ell) \propto \ell^{-1-\nu}$ so that only the
moments of order smaller than $\nu$ are finite; in particular, when
$\nu < 1$, the mean $\E\{\hat{\ell}\}$ is infinite, resulting in
anomalous properties such as a power-law decay of the probability
density of the reaction time in bounded domains (see
Sec. \ref{sec:surf_mech}).

(iii) If $\kappa(\ell)$ decreases faster than $1/\ell$, $K(\ell)$
approaches a finite limit $K(\infty)$ and thus $\Psi(\infty) =
e^{-K(\infty)} > 0$, i.e., the density $\psi(\ell)$ is not normalized
to $1$; as a consequence, the stopping local time $\hat{\ell}$ can be
infinite with a finite probability, $\P\{\hat{\ell} = \infty\} =
\Psi(\infty)$, i.e., the surface reaction may never occur in this case.  
This conclusion is rather expected: as the reactivity decreases fast,
the particle that did not react at first encounters, has less and less
chances to react at later ones.

\section{Generalized quantities}
\label{sec:generalized}

The alternative description developed in this work allows us to
incorporate various surface reaction mechanisms via an appropriate
stopping local time $\hat{\ell}$.  In fact, the average of the full
propagator $P(\x,\ell,t|\x_0)$ with the probability $\Psi(\ell) =
\P\{\hat{\ell} > \ell\}$ of no surface reaction up to $\ell$ yields a
generalized propagator that accounts for the survival against the
surface reaction determined by the stopping local time $\hat{\ell}$
(i.e., for the condition $\ell_t < \hat{\ell}$):
\begin{equation}
G_\psi(\x,t|\x_0) := \int\limits_0^\infty d\ell \, \Psi(\ell) \, P(\x,\ell,t|\x_0) .
\end{equation}
Substituting the spectral expansion (\ref{eq:Sigmap_spectral2}) for
the full propagator, we get in Laplace domain:
\begin{align}
& \tilde{G}_\psi(\x,p|\x_0) = \tilde{G}_\infty(\x,p|\x_0) \\   \nonumber
& \quad + \frac{1}{D} \sum\limits_{n} [V_n^{(p)}(\x_0)]^* \, V_n^{(p)}(\x) \, 
\int\limits_0^\infty d\ell\, \Psi(\ell) e^{-\mu_n^{(p)}\ell} \,,
\end{align}
where we used that $\Psi(0) = 1$.  Integrating by parts, one can also
write
\begin{align}  \label{eq:auxil11}
\tilde{G}_\psi(\x,p|\x_0) & = \tilde{G}_\infty(\x,p|\x_0) \\   \nonumber
& + \frac{1}{D} \sum\limits_{n} [V_n^{(p)}(\x_0)]^* \, V_n^{(p)}(\x) \, 
\frac{1 - \Upsilon_\psi(\mu_n^{(p)})}{\mu_n^{(p)}} \,,
\end{align}
where 
\begin{equation}  \label{eq:Upsilon}
\Upsilon_\psi(\mu) := \E\{ e^{-\mu \hat{\ell}} \} = \int\limits_0^\infty d\ell \, e^{-\mu \ell} \, \psi(\ell)
\end{equation}
is the Laplace transform of the probability density $\psi(\ell)$.  For
the exponential law of $\hat{\ell}$, one has $\Upsilon_\psi(\ell) =
1/(1 + \mu/q)$ and thus the conventional propagator admits the
following spectral expansion:
\begin{equation}
\tilde{G}_q(\x,p|\x_0) = \tilde{G}_\infty(\x,p|\x_0) + \frac{1}{D} \sum\limits_{n}  
\frac{[V_n^{(p)}(\x_0)]^* \, V_n^{(p)}(\x)}{q + \mu_n^{(p)}} \,.
\end{equation}
Setting $q = 0$, we obtain the identity
\begin{equation}
\frac{1}{D} \sum\limits_{n} \frac{[V_n^{(p)}(\x_0)]^* \, V_n^{(p)}(\x)}{\mu_n^{(p)}} = 
\tilde{G}_0(\x,p|\x_0) - \tilde{G}_\infty(\x,p|\x_0)  \,,
\end{equation}
that helps us to rewrite Eq. (\ref{eq:auxil11}) alternatively as
\begin{align} 
\tilde{G}_\psi(\x,p|\x_0) & = \tilde{G}_0(\x,p|\x_0) \\  \nonumber
& - \frac{1}{D} \sum\limits_{n} [V_n^{(p)}(\x_0)]^* \, V_n^{(p)}(\x) \, 
\frac{\Upsilon_\psi(\mu_n^{(p)})}{\mu_n^{(p)}} \,.
\end{align}

As previously, the normal derivative of the propagator defines the
probability flux density $j_\psi(\s,t|\x_0)$:
\begin{equation}  \label{eq:jpsi_Gpsi}
j_\psi(\s,t|\x_0) := - \left. D\bigl(\partial_n G_\psi(\x,t|\x_0)\bigr)\right|_{\x=\s}.
\end{equation}
The latter can also be expressed in terms of the surface hopping
propagator.  For this purpose, one can multiply the boundary condition
(\ref{eq:P_BC}) by $\Psi(\ell)$ and integrate over $\ell$ from $0$ to
$\infty$ to get
\begin{align*} 
\left. -D\bigl(\partial_n G_\psi(\x,t|\x_0)\bigr)\right|_{\x = \s} & = j_\infty(\s,t|\x_0)  \\
& + D\int\limits_0^\infty d\ell \, \Psi(\ell) \, \partial_\ell P(\s,\ell,t|\x_0)   ,
\end{align*}
where we used that $\Psi(0) = 1$.  The integration by parts yields
\begin{equation}  \label{eq:BC_aux2}
\left. -D\bigl(\partial_n G_\psi(\x,t|\x_0)\bigr)\right|_{\x = \s} = \int\limits_0^\infty d\ell \, \psi(\ell) \, D P(\s,\ell,t|\x_0)   ,
\end{equation}
where we used Eq. (\ref{eq:jinfty_P}) and the regularity condition
(\ref{eq:P_regul}).  We conclude thus
\begin{equation}  \label{eq:tildej_x0_3}
j_\psi(\s,t|\x_0) = \int\limits_0^\infty d\ell \, \psi(\ell) \, DP(\s,\ell,t|\x_0) .
\end{equation}

It is instructive to re-write Eq. (\ref{eq:BC_aux2}) as
\begin{equation}  \label{eq:cond_auxil22}
\int\limits_0^\infty d\ell \biggl(\Psi(\ell) \bigl(\partial_n P(\x,\ell,t|\x_0)\bigr)_{\x=\s} + \psi(\ell) \, P(\s,\ell,t|\x_0)\biggr) = 0.
\end{equation}
If $\psi(\ell) = q e^{-q\ell}$ (and thus $\Psi(\ell) = e^{-q\ell}$),
this identity implies the Robin boundary condition (\ref{eq:RobinA})
for the conventional propagator.  However, for other distributions of
the stopping local time $\hat{\ell}$, the generalized propagator
$G_\psi(\x,t|\x_0)$ does not satisfy the Robin boundary condition
(\ref{eq:RobinA}), i.e., $G_\psi(\s,t|\x_0)$ is not proportional to
$j_\psi(\s,t|\x_0)$.  In other words, setting the reactive flux
density in Eq. (\ref{eq:RobinA}) to be proportional to
$G_q(\s,t|\x_0)$ was a {\it choice}, one among many others.  This
choice represented the constant surface reactivity $\kappa$.
Expressing $\psi(\ell)$ and $\Psi(\ell)$ in terms of $\kappa(\ell)$
and $K(\ell)$ via Eq. (\ref{eq:psi_kappa}), one can also write
Eq. (\ref{eq:cond_auxil22}) as
\begin{align}  
& \int\limits_0^\infty d\ell \, e^{-K(\ell)} \left. \biggl( [D \partial_n + \kappa(\ell)] P(\x,\ell,t|\x_0)\biggr)\right|_{\x=\s} = 0 \, ,
\end{align}
which can be seen as an extension of the Robin boundary condition to
the case of encounter-dependent reactivity $\kappa(\ell)$.  We
emphasize that the integral form of this relation cannot be relaxed,
given that the full propagator $P(\x,\ell,t|\x_0)$ is unrelated to
surface reaction.

The Laplace transform of Eq. (\ref{eq:tildej_x0_3}) with respect to
$t$ yields 
\begin{equation}  \label{eq:tildej_x0}
\tilde{j}_\psi(\s,p|\x_0) = \int\limits_0^\infty d\ell \, \psi(\ell) \, D\tilde{P}(\s,\ell,p|\x_0) .
\end{equation}
This expression naturally extends Eq. (\ref{eq:identity}) to any
distribution of the stopping local time $\hat{\ell}$.  The spectral
expansion (\ref{eq:Sigmap_spectral}) implies
\begin{equation}  \label{eq:tildej_x0_1}
\tilde{j}_\psi(\s,p|\x_0) = \sum\limits_{n} [V_n^{(p)}(\s_0)]^* \, v_n^{(p)}(\s) \, \Upsilon_\psi(\mu_n^{(p)}),
\end{equation}
where $\Upsilon_\psi(\mu)$ is given by Eq. (\ref{eq:Upsilon}).

As in the conventional setting, the probability flux density
$j_\psi(\s,t|\x_0)$ is the joint probability density of the reaction
location $\X_\T$ and of the reaction time $\T$.  As a consequence, the
integral of $j_\psi(\s,t|\x_0)$ over $t$ gives the marginal
probability density of the reaction position (the so-called spread
harmonic measure in the conventional setting
\cite{Grebenkov06c,Grebenkov15}):
\begin{equation}
\omega_\psi(\s|\x_0) := \int\limits_0^\infty dt\, j_\psi(\s,t|\x_0) = \tilde{j}_\psi(\s,0|\x_0),
\end{equation}
whereas the integral of $j_\psi(\s,t|\x_0)$ over $\s$ yields the
marginal probability density of the reaction time,
\begin{equation}  \label{eq:Hpsi_def}
H_\psi(t|\x_0) := \int\limits_\pa d\s \, j_\psi(\s,t|\x_0).
\end{equation}
Substituting Eq. (\ref{eq:tildej_x0_1}), we get
\begin{align}  \nonumber
\omega_\psi(\s|\x_0) & = \int\limits_0^\infty d\ell \, \psi(\ell) \, D\tilde{P}(\s,\ell,0|\x_0) \\  \label{eq:omega_psi}
& = \sum\limits_{n}  [V_n^{(0)}(\x_0)]^* \, \Upsilon_\psi(\mu_n^{(0)}) \, v_n^{(0)}(\s)
\end{align}
and
\begin{equation} \label{eq:Hpsi}
\tilde{H}_\psi(p|\x_0) 
 = \sum\limits_{n} [V_n^{(p)}(\x_0)]^* \, \Upsilon_\psi(\mu_n^{(p)}) \int\limits_{\pa} d\s \, v_n^{(p)}(\s).
\end{equation}
The Laplace-transformed survival probability follows as
\begin{equation}
\tilde{S}_\psi(p|\x_0) = \frac{1-\tilde{H}_\psi(p|\x_0)}{p} \,.
\end{equation}

Substituting Eqs. (\ref{eq:Utilde_Sigma}, \ref{eq:tildej_x0}) into
Eq. (\ref{eq:Hpsi_def}), we get another representation
\begin{equation}  \label{eq:H_U1}
H_\psi(t|\x_0) = \int\limits_0^\infty d\ell \, \psi(\ell) \, U(\ell,t|\x_0) ,
\end{equation}
where $U(\ell,t|\x_0)$ is the probability density of the
first-crossing time.  One can see that $U(\ell,t|\x_0)$ plays a
crucial role in determining the properties of the reaction time for
{\it any surface reaction mechanism}, the latter being encoded via the
density $\psi(\ell)$ of the stopping local time $\hat{\ell}$.  As the
full propagator $P(\x,\ell,t|\x_0)$ is the fundamental intrinsic
characteristics of reflected Brownian motion, from which the
conventional propagator $G_q(\x,t|\x_0)$ follows via the Laplace
transform (\ref{eq:Pfull_Gq_SI}), the probability density
$U(\ell,t|\x_0)$ is the fundamental intrinsic characteristics of the
boundary local time, from which $H_\psi(t|\x_0)$ follows via
Eq. (\ref{eq:H_U1}).  In particular, the conventional density
$H_q(t|\x_0)$ is obtained for the exponential stopping local time:
\begin{equation}
H_q(t|\x_0) = \int\limits_0^\infty d\ell \, q e^{-q\ell} \, U(\ell,t|\x_0) ,
\end{equation}
from which one can also express $U(\ell,t|\x_0)$ as the inverse
Laplace transform with respect to $q$:
\begin{equation}
U(\ell,t|\x_0) = \L^{-1}_{\ell} \{ H_q(t|\x_0)/q\} .
\end{equation}

When there are many independent particles distributed with an initial
concentration $c(\x_0)$, their total diffusive flux, or the reaction
rate $J_\psi(t)$, can be obtained by averaging the probability fluxes
from all starting points $\x_0$ in the bulk:
\begin{equation}
J_\psi(t) := \int\limits_\Omega d\x_0 \, c(\x_0) \, H_\psi(t|\x_0).
\end{equation}
For a uniform concentration, $c(\x_0) = c_0$, we use the identity from
\cite{Grebenkov19}
\begin{equation}  \label{eq:identity_aux1}
\int\limits_\Omega d\x_0 \, V_n^{(p)}(\x_0) = \frac{D}{p} \mu_n^{(p)} \int\limits_{\pa} d\s \, v_n^{(p)}(\s) 
\end{equation}
to write
\begin{equation}  \label{eq:Jp_gen}
\tilde{J}_\psi(p) = \frac{c_0 D}{p} \sum\limits_{n} \left|\int\limits_{\pa} d\s \, v_n^{(p)}(\s)\right|^2 \, \mu_n^{(p)} \, 
\Upsilon_\psi(\mu_n^{(p)}).
\end{equation}
One can also represent the reaction rate as
\begin{equation}
J_\psi(t) = \int\limits_0^\infty d\ell \, \psi(\ell) \, U_{\rm uni}(\ell,t),
\end{equation}
where $U_{\rm uni}(\ell,t)$ is given by the inverse Laplace transform of
\begin{align} \nonumber
\tilde{U}_{\rm uni}(\ell,p) & := c_0 \int\limits_{\Omega} d\x_0 \, \tilde{U}(\ell,p|\x_0)  \\   \label{eq:U2_int}
& = \frac{c_0 D}{p} \sum\limits_n \mu_n^{(p)} \, e^{-\mu_n^{(p)} \ell} \left| \int\limits_{\pa} d\s \, v_n^{(p)}(\s) \right|^2  .
\end{align}
The above expressions couple the diffusive dynamics determined by the
eigenvalues and eigenfunctions of the Dirichlet-to-Neumann operator to
the surface reaction mechanism determined by $\Upsilon_\psi(\mu)$.  

We also mention that the symmetry of some domains can imply that the
ground eigenfunction $v_0^{(p)}(\s)$ is constant, whereas other
eigenfunctions should be orthogonal to it (see Sec. \ref{sec:shell}
for the example of a spherical shell).  In this special case, the
spectral expansions (\ref{eq:Hpsi}, \ref{eq:Jp_gen}) are reduced to a
single term depending on $\mu_0^{(p)}$, whereas a general surface
reaction mechanism encoded by the function $\Upsilon_\psi(\mu)$ is
formally equivalent to a partially reactive surface with a
``frequency''-dependent reactivity $\kappa_{\rm eff}(p)$ defined by
\begin{equation}
\frac{1}{1 + \mu_0^{(p)} D /\kappa_{\rm eff}(p)} = \Upsilon_\psi(\mu_0^{(p)}) \,.
\end{equation}
In time domain, such a reactivity corresponds to a convolution-type
Robin boundary condition.  While the above relation may provide an
additional insight onto our generalized surface reaction mechanisms,
this ``equivalence'' is very specific and valid only when a single
eigenmode contributes, e.g., for the probability density
$H_\psi(t|\x_0)$ and the associated reaction rate $J_\psi(t)$ in
selected symmetric domains.

\section{Spherical shell}
\label{sec:shell}

As an important practical example, we consider a spherical shell
between two concentric spheres of radii $R$ and $L$: $\Omega =
\{\x\in\R^3 ~:~ R < |\x| < L\}$.  While different combinations of
boundary conditions are possible, we focus on the typical case when
the reactive target of radius $R$ is confined by an outer reflecting
boundary of radius $L$.  The rotational invariance of $\Omega$ implies
that the eigenfunctions of the Dirichlet-to-Neumann operator, written
in spherical coordinates $(r,\theta,\phi)$, are the (normalized)
spherical harmonics (see \cite{Grebenkov19c,Grebenkov19} for details),
\begin{equation}  \label{eq:vn_ballI}
v_{nm}(\s) = \frac{1}{R} \, Y_{mn}(\theta,\phi)  \quad (n=0,1,2,\ldots, ~ |m| \leq n),
\end{equation}
which are independent of $p$.  In turn, the eigenvalues are
\begin{equation}  \label{eq:mu_shell}
\mu_n^{(p)} = - g'_n(R) ,
\end{equation}
where
\begin{equation}
g_n(r) = \frac{k'_n(\alpha L) i_n(\alpha r) - i'_n(\alpha L) k_n(\alpha r)}
{k'_n(\alpha L) i_n(\alpha R) - i'_n(\alpha L) k_n(\alpha R)} \,,
\end{equation}
$\alpha = \sqrt{p/D}$, prime denotes the derivative with respect to
the argument, and $i_n(z)$ and $k_n(z)$ are the modified spherical
Bessel functions of the first and second kind, respectively.  We
emphasize that the Dirichlet-to-Neumann operator is associated here to
the boundary local time exclusively on the inner sphere.  The
eigenvalue $\mu_n^{(p)}$ is $(2n+1)$ times degenerate.  In the limit
$L\to\infty$, one gets
\begin{equation}   \label{eq:mu_ballE}
\mu_n^{(p)} = - \sqrt{p/D} \, \frac{k'_n(R\sqrt{p/D})}{k_n(R\sqrt{p/D})}
\end{equation}
for the exterior of a ball.  In the limit $p\to 0$, one has
\begin{equation}
\mu_n^{(0)} = \frac{n+1}{R} \, \frac{1 - (R/L)^{2n+1}}{1 + \frac{n+1}{n} (R/L)^{2n+1}} \,,
\end{equation}
which reduces to $(n+1)/R$ as $L\to\infty$.

One also needs to compute $V_n^{(p)}(\x_0)$ from Eq. (\ref{eq:Vnp}).
Using the summation formulas from \cite{Grebenkov19g}, the
Laplace-transformed propagator $\tilde{G}_\infty(\x,p|\x_0)$ and thus
$\tilde{j}_\infty(\s,p|\x_0)$ for a spherical shell with Dirichlet
boundary condition on the inner sphere and Neumann boundary condition
on the outer sphere read
\begin{align} \nonumber
\tilde{G}_\infty(\x,p|\x_0) &= \sum\limits_{n=0}^\infty \frac{\alpha (2n+1)}{4\pi D}  P_n\biggl(\frac{(\x \cdot \x_0)}{|\x| \, |\x_0|}\biggr)
g_n(r_0)  \\
& \times \bigl[k_n(\alpha R) i_n(\alpha r) - i_n(\alpha R) k_n(\alpha r)\bigr] ,\\
\tilde{j}_\infty(\s,p|\x_0) &= \sum\limits_{n=0}^\infty \frac{2n+1}{4\pi R^2} \, P_n\biggl(\frac{(\s \cdot \x_0)}{|\s| \, |\x_0|}\biggr)
g_n(r_0) ,
\end{align}
where $r = |\x|$, $r_0 = |\x_0|$, $R \leq r \leq r_0 \leq L$, $P_n(z)$
are the Legendre polynomials, and we used the Wronskian $i'_n(z)
k_n(z) - k'_n(z) i_n(z) = 1/z^2$.  The projection of
$\tilde{j}_\infty(\s,p|\x_0)$ onto an eigenfunction $v_{nm}(\s)$ from
Eq. (\ref{eq:vn_ballI}) reads then
\begin{equation}
V_{nm}^{(p)}(\x_0) = v_{mn}(\theta_0,\phi_0) \, g_n(r_0) .
\end{equation}
The orthogonality of spherical harmonics reduces Eq. (\ref{eq:U2}) to
\begin{equation}   \label{eq:U_shell}
U(\ell,t|\x_0) = \L^{-1}_t\bigl\{ g_0(r_0) \, \exp(-\mu_0^{(p)}\ell) \bigr\} ,
\end{equation}
while the probability density of the reaction time reads
\begin{equation}   \label{eq:Hpsi_shell}
H_\psi(t|\x_0) = \L^{-1}_t \bigl\{ g_0(r_0) \, \Upsilon_\psi(\mu_0^{(p)}) \bigr\} \,,
\end{equation}
where
\begin{align*}
\mu_0^{(p)} & = (\alpha + 1/R) \frac{1 - v(R) \frac{e^{-2\alpha R} + (\alpha R - 1)/(\alpha R+1)}{1 - e^{-2\alpha R}}}{1 + v(R)} \,, \\
g_0(r_0) & = \frac{R}{r_0} e^{-\alpha(r_0-R)} \frac{1 + v(r_0)}{1 + v(R)} \,, \\
v(r) & = e^{-2\alpha(L-r)} \frac{1 - e^{-2\alpha r}}{e^{-2\alpha L} + \frac{\alpha L-1}{\alpha L+1}} \,.
\end{align*}
Similarly, Eq. (\ref{eq:U2_int}) reads
\begin{equation}
U_{\rm uni}(\ell,t) = 4\pi R^2 c_0 D \, \L^{-1}_t \biggl\{ \frac{\mu_0^{(p)}}{p} \, \exp(-\mu_0^{(p)}\ell) \biggr\}.
\end{equation}

\section{Surface reaction models}
\label{sec:surf_mech}

\begin{table*}
\centering
\begin{tabular}{|c|c|c|c|c|c|c|l|} \hline
n. & Distribution & $\psi(\ell)$ & $\Upsilon_\psi(\mu)$ & $\kappa(\ell)/\kappa_0$ & $\ell\to 0$ & $\ell\to\infty$ & Comment \\ \hline
1 & exponential & $q e^{-q\ell}$ & $(1+\mu/q)^{-1}$ & $1$ & $1$ & $1$ & \\  \hline
2 & gamma       & $\displaystyle q e^{-q \ell}  \,\frac{(q\ell)^{\nu-1} }{\Gamma(\nu)}$ & $(1 + \mu/q)^{-\nu}$ & 
	$\displaystyle \frac{(q\ell)^{\nu-1} e^{-q\ell}}{\Gamma(\nu,q\ell)}$ & $\displaystyle \frac{(q\ell)^{\nu-1}}{\Gamma(\nu)}$ 
	& $1$ & $\nu > 0$  \\  \hline
3 & \parbox{15mm}{Pareto II (Lomax)} & $q \nu (1 + q\ell)^{-1-\nu}$ & $\nu (\mu/q)^{\nu} \, e^{\mu/q} \Gamma(-\nu, \mu/q)$ 
	& $\nu/(1 + q\ell)$ & $\nu$ & $\nu/(q\ell)$ & $\nu > 0$ \\    \hline
4 & Mittag-Leffler & $-E_{\nu,0}(-(q\ell)^\nu)/\ell$ & $1/(1 + (\mu/q)^\nu)$ & 
	$\displaystyle (q\ell)^{\nu-1} \frac{E_{\nu,\nu}(-(q\ell)^\nu)}{E_{\nu,1}(-(q\ell)^\nu)}$ & 
	$\displaystyle \frac{(q\ell)^{\nu-1}}{\Gamma(\nu)}$ & $\nu/(q\ell)$ & $0 < \nu < 1$ \\  \hline
\multirow{2}{2mm}{5} & \multirow{2}{18mm}{\parbox{18mm}{power-law reactivity}}
	& \multicolumn{2}{l|}{$q\beta (1+q\ell)^\nu \exp\bigl(\beta \frac{1-(1+q\ell)^{\nu+1}}{\nu+1}\bigr)$ \hskip 6mm (see caption)}  
	& $\beta (1 + q\ell)^\nu$ & $\beta$ & $\beta (q\ell)^{\nu}$ & $\nu \ne -1$ \\  \cline{3-8}
  &    & \multicolumn{2}{l|}{$q \beta(q\ell)^\nu \exp\bigl(-\beta\frac{(q\ell)^{\nu+1}}{\nu+1}\bigr)$ \hskip 16mm (see caption)}
	& $\beta(q\ell)^\nu$ & $\beta(q\ell)^\nu$ & $\beta(q\ell)^\nu$ & $\nu > -1$ \\  \hline
6 & \parbox{18mm}{exponential reactivity}    
	& $q\nu \exp(-q\ell-\nu(1-e^{-q\ell}))$ & $\displaystyle \frac{\nu M(\mu/q+1,\mu/q+2;\nu)}{e^{\nu} (\mu/q+1)}$ & 
	$\nu \exp(-q\ell)$ & $\nu$ & $\nu e^{-q\ell}$ &  \\  \hline
7 & \parbox{18mm}{truncated exponential} & $q e^{-q(\ell-\ell_1)} {\I}_{(\ell_1,\ell_2)}(\ell)$ & 
	$\displaystyle e^{-\mu\ell_1}\frac{1-e^{-(\mu + q)(\ell_2-\ell_1)}}{1 + \mu/q}$ 
	& $\I_{(\ell_1,\ell_2)}(\ell)$ & $0$ or $1$ & $0$ or $1$ & $\ell_1 < \ell_2$ \\ \hline 
8 & L\'evy-Smirnov & $\displaystyle q \frac{\exp(-1/(q\ell))}{\sqrt{\pi} \, (q\ell)^{3/2}}$ & $\exp(-2\sqrt{\mu/q})$ & 
	$\displaystyle \frac{\exp(-1/(q\ell))}{\sqrt{\pi} (q\ell)^{3/2} \erf(\sqrt{1/(q\ell)})}$ & 
	$\sim e^{-1/(q\ell)}$ & $1/(2q\ell)$ &  \\ \hline
9 & \parbox{15mm}{one-sided Gaussian} & $2q \, e^{-(q\ell)^2}/\sqrt{\pi}$ & $\erfcx(\mu/(2q))$ & $\displaystyle \frac{2}{\sqrt{\pi} \, \erfcx(q\ell)}$ 
	& $2/\sqrt{\pi}$ & $2q\ell$ &  \\ \hline 
\end{tabular}
\caption{
Selected surface reaction models.  The 2nd, 3rd and 4th columns
present the probability density $\psi(\ell)$, its Laplace transform
$\Upsilon_\psi(\mu)$, and the corresponding encounter-dependent
reactivity $\kappa(\ell)/\kappa_0$, rescaled by some reference
reactivity $\kappa_0$ such that $q = \kappa_0/D$.  The 5th and 6th
columns show the leading term of the asymptotic behavior of
$\kappa(\ell)/\kappa_0$ at small and large $\ell$, respectively.  Here
$\Gamma(\nu,z)$ is the upper incomplete gamma function; $\erfcx(z) =
e^{z^2} \erfc(z)$ is the scaled complementary error function;
$E_{\alpha,\beta}(z)$ is the Mittag-Leffler function; $M(a,b;z) =
\,_1F_1(a,b;z)$ is the Kummer's confluent hypergeometric function;
$\I_{(a,b)}(x)$ is the indicator function ($\I_{(a,b)}(x) = 1$ if
$x\in(a,b)$ and $0$ otherwise).  For the model 5 (top line),
$\Upsilon_\psi(\mu) = \alpha \beta\, e^{\alpha\beta+\mu/q}
\int\nolimits_1^\infty dy \, e^{-\alpha \beta y - (\mu/q) y^\alpha}$
for $\alpha > 0$ and $\Upsilon_\psi(\mu) = |\alpha| \beta \,
e^{\alpha\beta +\mu/q} \int\nolimits_0^1 dy \, e^{-\alpha \beta y -
(\mu/q) y^\alpha}$ for $\alpha < 0$, with $\alpha = 1/(\nu+1)$.  For
the model 5 (bottom line), $\Upsilon_\psi(\mu) =
\alpha \beta \int\nolimits_0^\infty dy \, e^{-\alpha \beta y - (\mu/q)
y^\alpha}$.  For the Pareto-II model with $\nu = 1/2$, one can write
$\Upsilon_\psi(\mu) = 1 - \sqrt{\pi \mu/q}\, \erfcx(\sqrt{\mu/q})$. }
\label{tab:psi}
\end{table*}

In this section, we discuss several models of surface reaction
mechanisms, which are determined by choosing an appropriate
distribution for the stopping local time $\hat{\ell}$.  One can select
either the probability density $\psi(\ell)$ capturing the desired
stopping criterion, or the desired dependence of the reactivity
$\kappa(\ell)$ on the local time $\ell$.  According to
Eqs. (\ref{eq:psi_kappa}, \ref{eq:kappa_psi}), both options are
equivalent, i.e., $\psi(\ell)$ determines $\kappa(\ell)$, and
vice-versa.

\subsection{Selected models}

\begin{figure}
\begin{center}
\includegraphics[width=42mm]{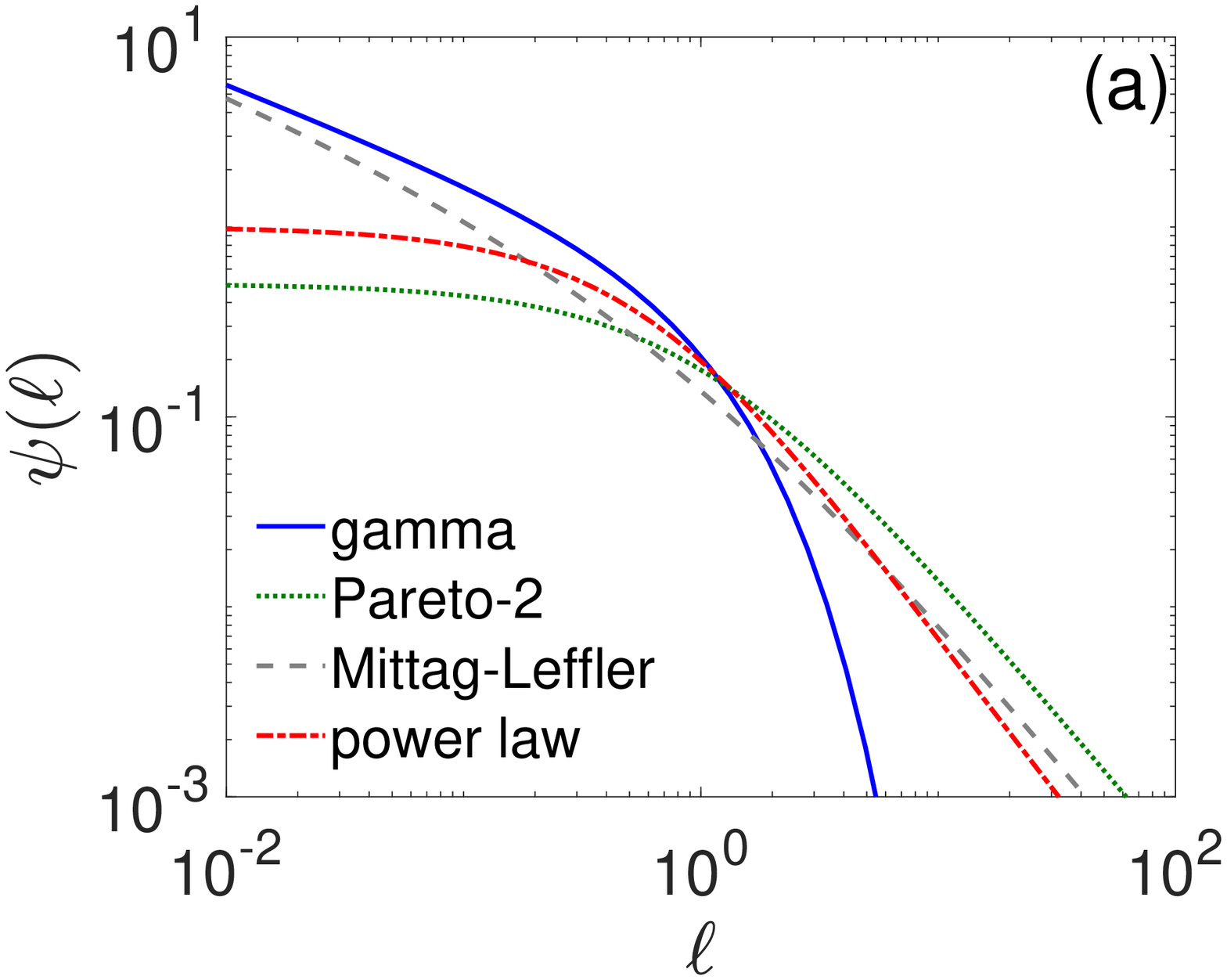} 
\includegraphics[width=42mm]{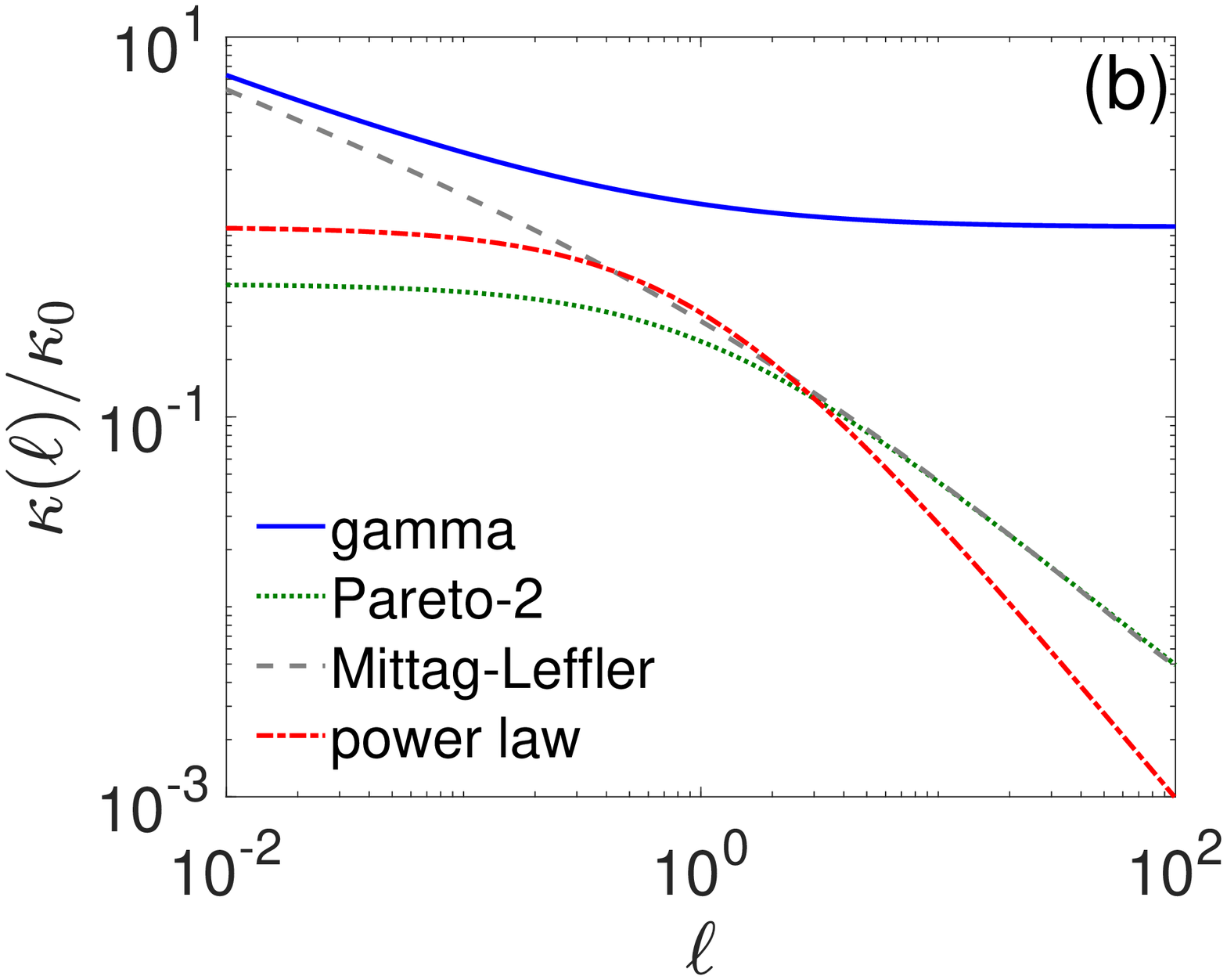} 
\includegraphics[width=42mm]{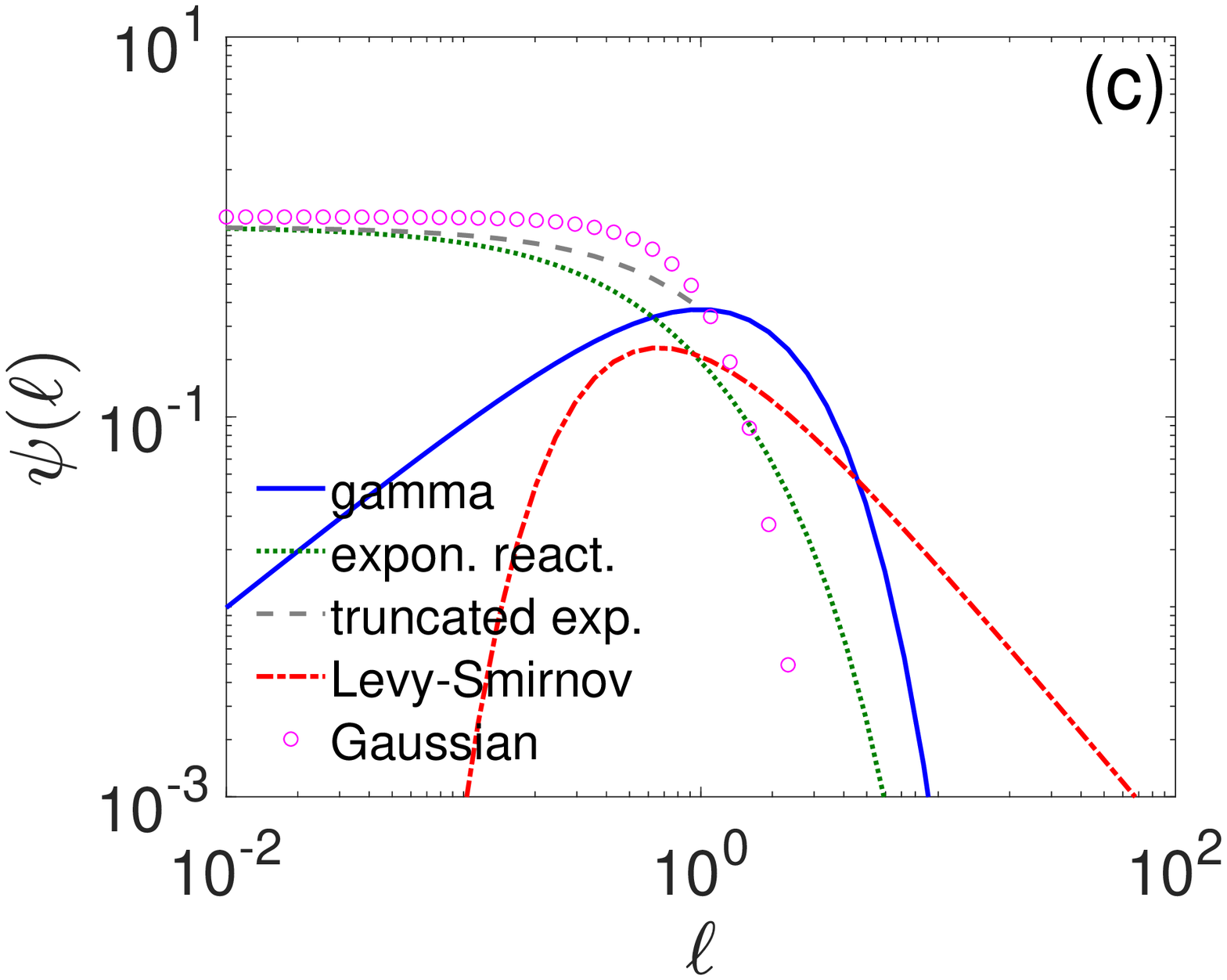} 
\includegraphics[width=42mm]{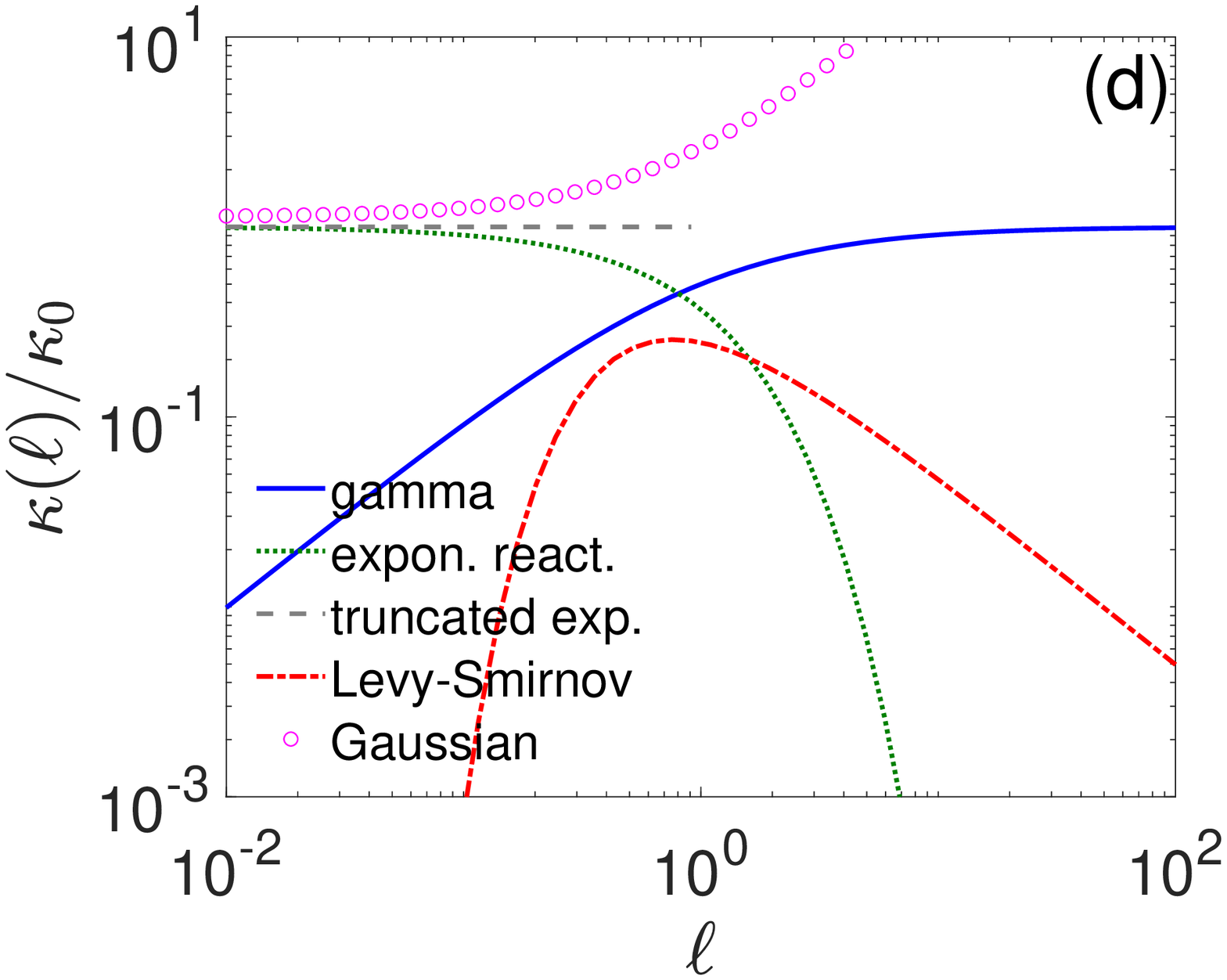} 
\end{center}
\caption{
Selected surface reaction models: {\bf (a,c)} Probability density
$\psi(\ell)$ of the stopping local time $\hat{\ell}$; {\bf (b,d)}
Encounter-dependent reactivity $\kappa(\ell)/\kappa_0$.  Panels {\bf
(a,b)} present the gamma model ($\nu = 0.5$), the Pareto-II model
($\nu = 0.5$), the Mittag-Leffler model ($\nu = 0.5$) and the
power-law reactivity model ($\nu = -1.5$).  Panels {\bf (c,d)} present
the gamma model ($\nu = 2$), the exponentially decaying reactivity
model ($\nu = 1$), the truncated exponential model ($\ell_1 = 0$,
$\ell_2 = 1$), the L\'evy-Smirnov model, and the one-sided Gaussian
model.  In all cases, $q = 1$.}
\label{fig:models}
\end{figure}

Table \ref{tab:psi} lists selected distributions for the stopping
local time $\hat{\ell}$, which can be used to produce a variety of
surface reaction mechanisms (see also Fig. \ref{fig:models}).
Expectedly, only the exponential distribution of $\hat{\ell}$ yields
Poissonian-like reactions with a constant reactivity.  Using the gamma
distribution with the scale parameter $\nu$, one can either increase
($\nu < 1$) or diminish ($\nu > 1$) the probability of small values of
$\hat{\ell}$ and thus control the reactivity $\kappa(\ell)$ at small
local times.  For $\nu > 1$, one can model reaction mechanisms, in
which the surface is initially inactive and needs to be activated
(``heated up'') by repeated encounters with the particle to start
working with a constant reactivity.  Alternatively, the gamma
distribution with $\nu < 1$ may describe situations when the surface
is highly reactive at the beginning and then, after a number of
encounters with the particle, reaches a lower constant reactivity.

In turn, a Pareto-type (or Lomax) distribution keeps the reactivity
$\kappa(\ell)$ constant at small $\ell$ but then leads to a power-law
decay as $1/\ell$ at large $\ell$.  Such a distribution can model slow
passivation (or aging) of the reactive surface due to repeated
encounters with the particle.  This example also illustrates that
different densities $\psi(\ell)$ (with different exponent $\nu$) can
result in a very similar behavior of the encounter-dependent
reactivity $\kappa(\ell)$ (which differ here only by a prefactor
$\nu$).  Note that Mittag-Leffler distribution allows one to represent
the decrease of reactivity even at small values of $\ell$.

More generally, Eq. (\ref{eq:psi_kappa}) helps one to incorporate a
faster decay of the reactivity at large $\ell$ such as a power law
$\ell^{-\nu}$ with $\nu > 1$, an exponential $e^{-q\ell}$, or any
desired behavior.  In this case, the probability density $\psi(\ell)$
is not normalized to $1$, and the surface reaction does not occur with
the probability $e^{-K(\infty)}$ (see Sec. \ref{sec:passivation}).
This is particularly clear for the truncated exponential distribution,
for which the surface becomes totally inert after a number of
encounters (when $\ell > \ell_2$).

While previous examples provided monotonous behavior of the
reactivity, the L\'evy-Smirnov distribution results in a maximum of
$\kappa(\ell)$.  In this setting, the surface is almost inert for
$\ell \ll 1/q$, becomes most reactive at $\ell_c \approx 0.7667/q$,
and then slowly looses its reactivity.  This model can describe
surfaces that have an optimal range of reactivity.

Finally, the reactivity $\kappa(\ell)$ can also be increasing with
$\ell$.  For instance, a linear asymptotic increase can be modeled by
using one-sided Gaussian distribution of $\hat{\ell}$.  Note that a
faster increase of the reactivity implies a faster decrease of
$\psi(\ell)$ at large $\ell$.

\subsection{Reaction rate on a spherical target}

Our probabilistic description couples the surface reaction mechanism
to diffusive dynamics.  In order to illustrate the effect of various
mechanisms, we consider an archetypical setting in diffusion-mediated
chemical kinetics -- a spherical target of radius $R$ in the
three-dimensional space.  For this target, Smoluchowski first derived
the reaction rate for perfect reactions \cite{Smoluchowski17} and
later Collins and Kimball extended it to partial reactivity
\cite{Collins49}.

The eigenvalues and eigenfunctions of the Dirichlet-to-Neumann
operator for this domain are summarized in Sec. \ref{sec:shell}.  The
orthogonality of spherical harmonics removes all the terms in
Eq. (\ref{eq:Jp_gen}), except the first one that gives
\begin{equation}  \label{eq:Jp_ball}
\tilde{J}_\psi(p) = J_{\rm S} \, \frac{R}{p} \mu_0^{(p)} \,  \Upsilon_\psi(\mu_0^{(p)}),
\end{equation}
where $\mu_0^{(p)} = \sqrt{p/D} + 1/R$ from Eq. (\ref{eq:mu_ballE}),
and
\begin{equation}
J_{\rm S} = 4\pi R c_0 D
\end{equation}
is the steady-state Smoluchowski rate.  For the conventional
Poissonian-like surface reaction, one has $\Upsilon(\mu) = 1/(1 +
\mu/q)$, and the Laplace transform inversion of Eq. (\ref{eq:Jp_ball})
yields the Collins-Kimball's rate \cite{Collins49}:
\begin{equation}   \label{eq:Jt_CK}
\frac{J_q(t)}{J_{\rm S}} = \frac{1}{1 + 1/(qR)} \biggl\{1 + qR \, \erfcx\biggl(\sqrt{Dt}\, (q + 1/R)\biggr)\biggr\} ,
\end{equation}
where $\erfcx(z) = e^{z^2} \erfc(z)$ is the scaled complementary error
function.  The function $J_q(t)$ varies from the reaction-limited rate
$4\pi R^2 c_0 \kappa$ at $t = 0$ to the diffusion-limited rate $4\pi R
c_0 D/(1 + 1/(qR))$ as $t\to\infty$.  The limit $q\to \infty$ gives
the Smoluchowski's rate \cite{Smoluchowski17}:
\begin{equation}  \label{eq:Jt_Smol}
\frac{J_\infty(t)}{J_{\rm S}} =  1 + \frac{R}{\sqrt{\pi Dt}} \, .
\end{equation}
For a general surface reaction mechanism, we perform the inverse
Laplace transform of Eq. (\ref{eq:Jp_ball}) numerically by the Talbot
algorithm.

The long-time behavior of the reaction rate, which is determined by
the limit $p\to 0$, is universal.  In fact, as $\mu_0^{(p)}$
approaches a strictly positive constant $\mu_0^{(0)} = 1/R$ in this
limit, the reaction rate $J_\psi(t)$ approaches its steady-state limit:
\begin{equation}
\frac{J_\psi(\infty)}{J_{\rm S}} = \Upsilon_\psi(1/R) \leq 1.
\end{equation}
In other words, all surface reaction mechanisms lead to a constant
steady-state limit at long times, while the function
$\Upsilon_\psi(\mu)$ determines its level.  We emphasize that this is
a general property for the exterior of any compact domain in $\R^3$,
for which $\mu_0^{(0)} > 0$ (even though in general, other terms in
Eq. (\ref{eq:Jp_gen}) would contribute).  In contrast, the way how
$J_\psi(t)$ reaches this limit, as well as the short-time behavior,
are not universal.

\begin{figure}
\begin{center}
\includegraphics[width=88mm]{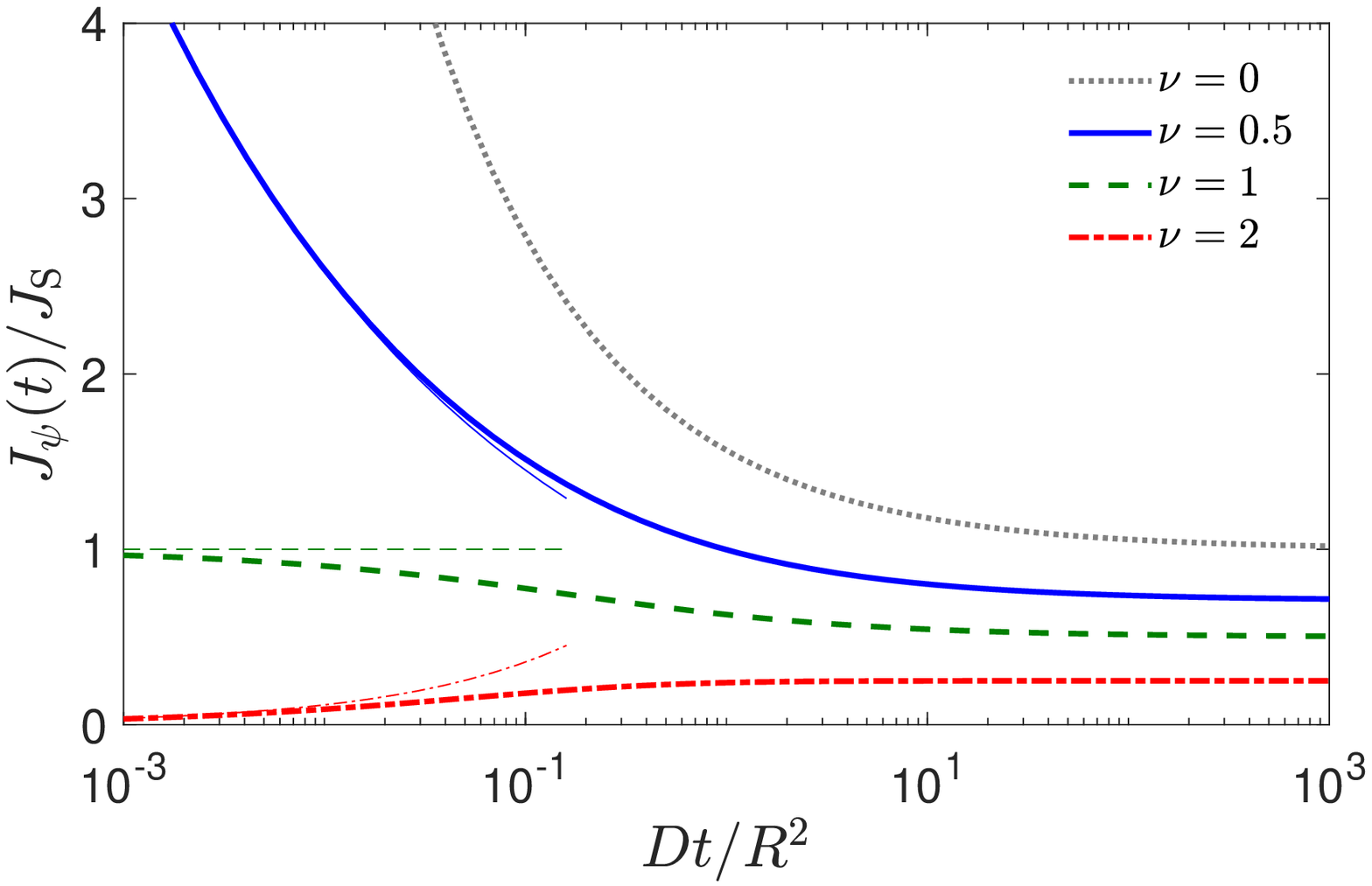} 
\end{center}
\caption{
The reaction rate $J_\psi(t)$, rescaled by the Smoluchowski
steady-state rate $J_{\rm S} = 4\pi R c_0 D$, on a spherical target of
radius $R$, for the gamma distribution of the stopping local time
$\hat{\ell}$: $\psi(\ell) = q(q\ell)^{\nu-1} e^{-q\ell}/\Gamma(\nu)$,
with $qR = 1$ and three values of $\nu$: $\nu = 0.5$ (solid line),
$\nu = 1$ (dashed line, Collins-Kimball rate $J_q(t)$ from
Eq. (\ref{eq:Jt_CK})), $\nu = 2$ (dash-dotted line).  The associated
thin lines present the short-time power law (\ref{eq:Jt_gamma_t0}) for
these three cases.  Dotted line shows the Smoluchowski rate
$J_\infty(t)$ from Eq. (\ref{eq:Jt_Smol}) for a perfectly reactive
target (limit $\nu = 0$).}
\label{fig:Jt_gamma}
\end{figure}

We illustrate the effect of encounter-dependent reactivity onto the
reaction rate by considering the gamma model (see Table
\ref{tab:psi}) which naturally generalizes the conventional
exponential distribution.  Figure \ref{fig:Jt_gamma} shows the
reaction rate $J_\psi(t)$ for a weakly reactive target with $q R = 1$.
As discussed above, the shape parameter $\nu$ determines the behavior
of the reactivity $\kappa(\ell)$ at small $\ell$ and thus strongly
affects the reaction rate at short times.  Indeed, the short-time
behavior of $J_\psi(t)$ can be deduced from the asymptotic analysis of
$\tilde{J}_\psi(p)$ as $p\to\infty$:
\begin{equation}  \label{eq:Jt_gamma_t0}
\frac{J_\psi(t)}{J_S} = \frac{R q^\nu}{\Gamma\bigl(\frac{\nu+1}{2}\bigr)} \, (Dt)^{(\nu-1)/2} + O(t^{\nu/2}) .
\end{equation}
When $\nu > 1$, the reactivity $\kappa(\ell)$ (see Table
\ref{tab:psi}) vanishes at small $\ell$ so that the diffusing particle
has low chances to react on the target at first encounters, and the
reaction rate is accordingly small at short times.  Only after a
number of returns to the target, the reaction becomes probable that
can model situations when the reactive surface needs to be
progressively activated (``heated up'') by repeated encounters with
the diffusing particle.  The value $\nu = 1$ corresponds to the
conventional constant reactivity so that one retrieves the
Collins-Kimball result (\ref{eq:Jt_CK}).  In contrast, for $0 < \nu <
1$, the high reactivity at small $\ell$ facilitates surface reactions
at short times, enhancing the reaction rate.  This setting can model
surfaces which are highly reactive at the beginning and then reach a
lower steady-state reactivity.  Note also that in the limit $\nu \to
0$, the gamma distribution degenerates into a Dirac distribution
$\delta(\ell)$ so that this limit corresponds to a perfectly absorbing
boundary, on which the reaction occurs at the first encounter.  The
reaction rate $J_\psi(t)$ approaches thus $J_\infty(t)$ from
Eq. (\ref{eq:Jt_Smol}).

\subsection{Distribution of the reaction time}
\label{sec:distrib_sphere}

The distribution of the reaction time $\T$ on a target provides finer
information about diffusion-influenced reactions.  To go beyond the
basic example of a spherical target in the whole space $\R^3$, we
consider now a spherical target of radius $R$ surrounded by an outer
reflecting concentric sphere of radius $L$ that confines diffusing
particles within a bounded region around the target.  In other words,
we consider a spherical shell between two concentric spheres of radii
$R$ and $L$ (see Sec. \ref{sec:shell}).

\begin{figure}
\begin{center}
\includegraphics[width=88mm]{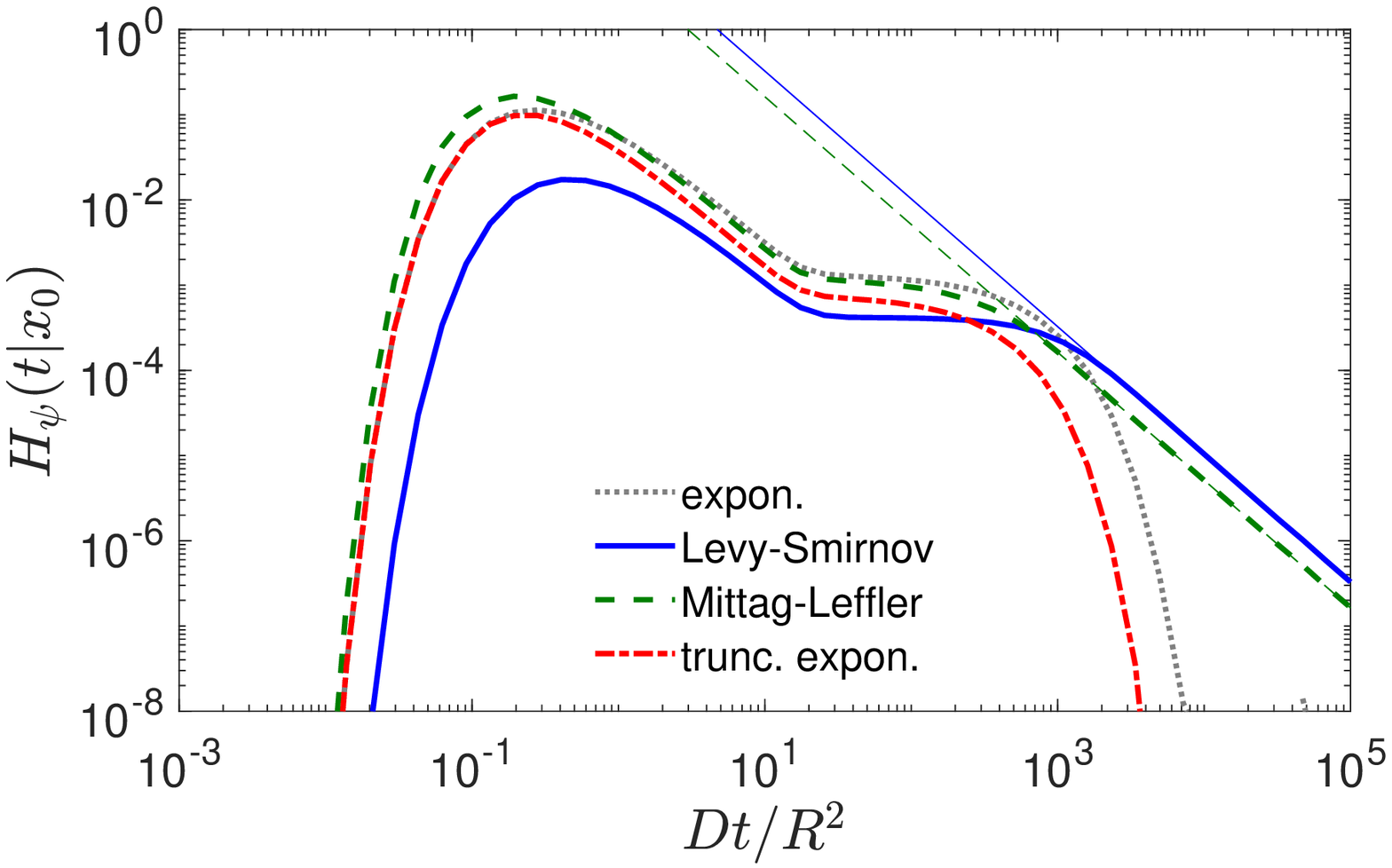} 
\end{center}
\caption{
The probability density $H(t|\x_0)$ (rescaled by $R^2/D$) of the
reaction time on a spherical target of radius $R$, surrounded by an
outer reflecting concentric sphere of radius $L = 10R$, for several
surface reaction models and the starting point $|\x_0| = 2R$.  Four
models with $qR = 1$ are compared: the conventional exponential model
(dotted line), the L\'evy-Smirnov model (solid line), the
Mittag-Leffler model with $\nu = 1/2$ (dashed line), and the truncated
exponential model with $\ell_1 = 0$ and $\ell_2 = 0.5 R$ (dash-dotted
line).  Thin lines show the long-time asymptotic behavior
(\ref{eq:Hpsi_asympt0}) for the LS and ML models.}
\label{fig:Ht_comparison}
\end{figure}

The distribution of the reaction time on a partially reactive target
surrounded by an outer reflecting sphere has been thoroughly
investigated for conventional Poissonian surface reaction (see
\cite{Grebenkov18} and references therein).  Figure
\ref{fig:Ht_comparison} presents the probability density $H_\psi(t|\x_0)$
for several models of the surface reaction mechanism.  Here we fix
$|\x_0| = 2R$, $L = 10R$ and $qR = 1$.  Apart from the conventional
exponential model, we consider the L\'evy-Smirnov (LS) model, the
Mittag-Leffler (ML) model, and the truncated exponential (TE) model
with $\ell_1 = 0$ and $\ell_2 = 0.5 R$.  In all cases,
$H_\psi(t|\x_0)$ is computed numerically via the inverse Laplace
transform of Eq. (\ref{eq:Hpsi_shell}) evaluated by using the Talbot
algorithm.

At short times, the probability density $H_\psi(t|\x_0)$ for the ML
and TE models is close to that of the exponential model.  In fact,
this behavior describes rare particles that move directly towards the
target and rapidly react.  Even though the reactivity $\kappa(\ell)$
is increasing for the ML model, the overall reaction is still
determined by diffusion.  This situation is drastically different for
the LS model, for which the reactivity $\kappa(\ell)$ is extremely
small at first encounters, thus shifting the probability density
toward longer times.  In turn, the LS and ML models, for which
$\psi(\ell)$ has the same power law decay as $\ell\to\infty$
($\psi(\ell) \propto \ell^{-3/2}$), exhibit similar long-time
behavior:
\begin{equation}   \label{eq:Hpsi_asympt0}
H_\psi(t|\x_0) \propto t^{-3/2}  \qquad (t\to\infty).
\end{equation}
The power-law decay of $H_\psi(t|\x_0)$ for a bounded domain is a new
feature of the considered surface reaction mechanisms.  The decreasing
reactivity $\kappa(\ell)$ facilitates the survival of particles even
in bounded domains.  A similar effect was earlier reported for
continuous-time random walks in bounded domains
\cite{Grebenkov10b}, but it was related to long stalling periods that
a particle experiences during its motion to the target.  In our
setting, the particle diffuses normally, and the enhancement of the
survival probability is caused by the surface reaction mechanism.  The
long-time asymptotic formula (\ref{eq:Hpsi_asympt0}) is in excellent
agreement with $H_\psi(t|\x_0)$.
Finally, in the case of the truncated exponential model, the
probability density $H_\psi(t|\x_0)$ decays exponentially again but it
is not normalized to $1$ anymore.  In fact, as the boundary becomes
inert for $\ell > \ell_2$, the particle that failed to react up to the
boundary local time $\ell_2$, survives forever.  In this example,
$\P\{\T = \infty\} = \exp(-q\ell_2) \approx 0.61$.

We emphasize the generic character of the power-law decay of
$H_\psi(t|\x_0)$ in bounded domains in the critical regime when
$\kappa(\ell) \simeq \nu D/\ell$ at large $\ell$ (with some
dimensionless constant $0 < \nu < 1$).  In fact, the long-time
asymptotic behavior of $H_\psi(t|\x_0)$ is determined by the small-$p$
behavior of $\tilde{H}_\psi(p|\x_0)$ and thus of
$\tilde{U}(\ell,p|\x_0)$, due to Eq. (\ref{eq:H_U1}).  For a bounded
domain $\Omega$, the latter behavior is determined by the ground
eigenmode, for which $\mu_0^{(p)} \simeq \gamma p + O(p^2)$ as $p\to
0$, with $\gamma = |\Omega|/(D|\pa|)$ \cite{Grebenkov19c}.  As a
consequence, we have
\begin{equation}  \label{eq:U_auxil2}
\tilde{U}(\ell,p|\x_0) \simeq \exp(-\gamma p\ell)   \qquad (p\to 0), 
\end{equation}
where we used that $v_0^{(p)}(\s) \to |\pa|^{-1/2}$ as $p\to 0$, the
orthogonality of other eigenfunctions $v_n^{(p)}(\s)$ to
$v_0^{(p)}(\s)$, and that $V_0^{(p)}(\x_0) \to |\pa|^{-1/2}$.
Integrating the expression (\ref{eq:U_auxil2}) with $\psi(\ell)
\propto \ell^{-1-\nu}$ (see Sec. \ref{sec:passivation}), one gets
$\tilde{H}_\psi(p|\x_0) \propto p^{\nu}$ as $p\to 0$, from which
\begin{equation}  \label{eq:Hpsi_asympt}
H_\psi(t|\x_0) \propto t^{-1-\nu}  \qquad (t\to \infty).
\end{equation}
The same asymptotic behavior holds for the reaction rate: $J_\psi(t)
\propto t^{-1-\nu}$.  Such a long-time asymptotic decay of the
reaction time probability density is unusual for bounded domains, for
which the conventional constant reactivity implies the exponential
decay.

\begin{figure}
\begin{center}
\includegraphics[width=88mm]{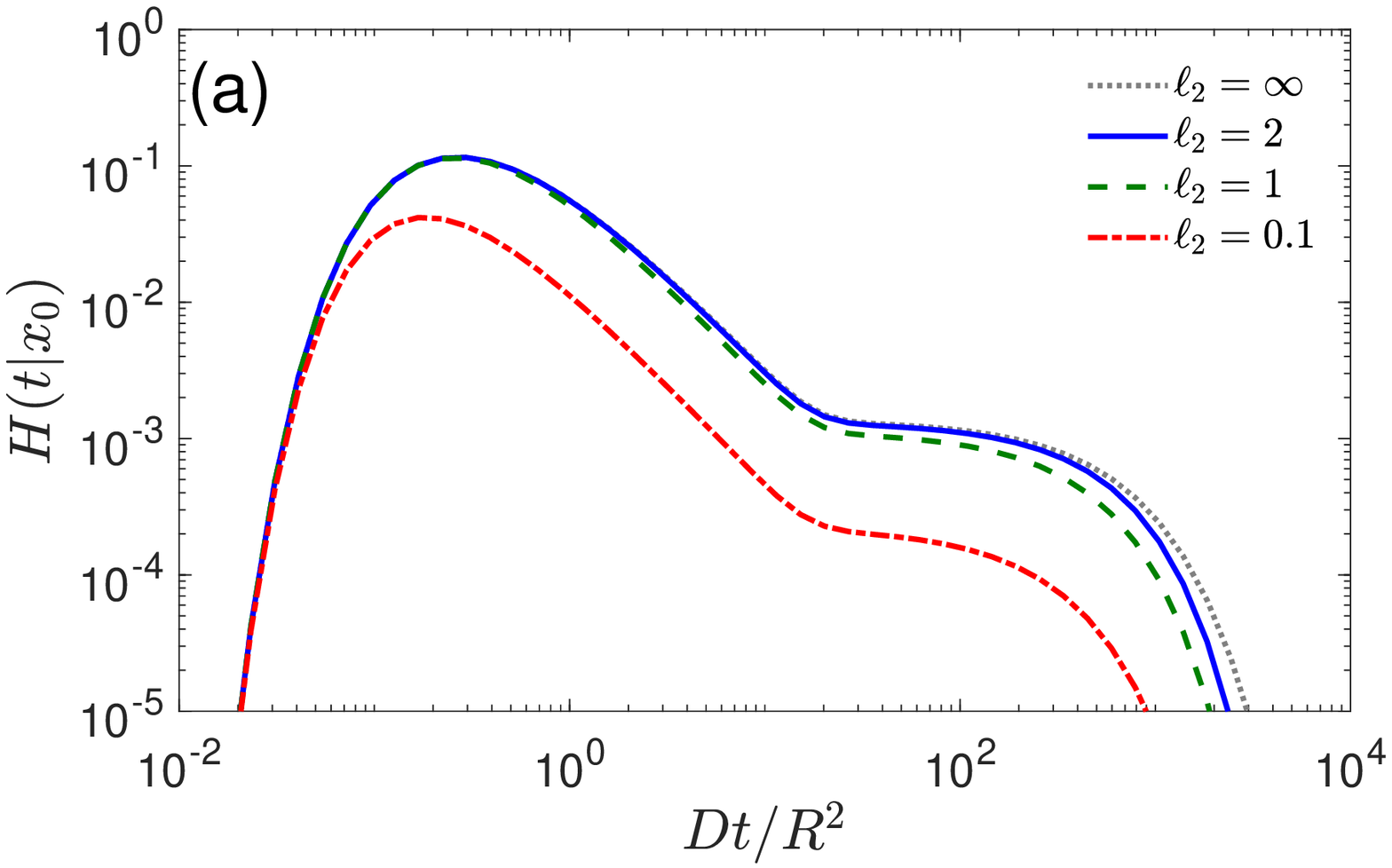} 
\includegraphics[width=88mm]{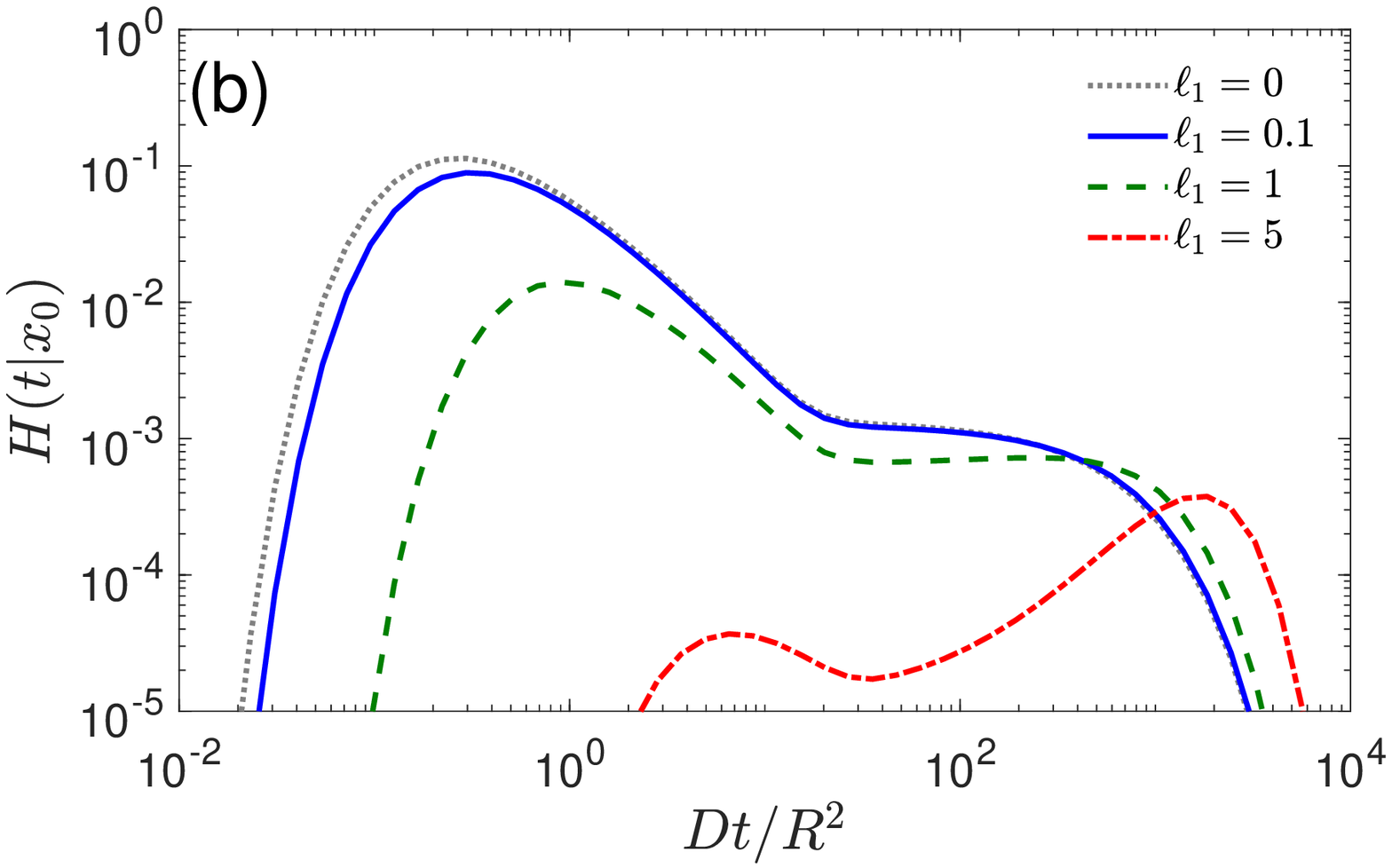} 
\end{center}
\caption{
The probability density $H_\psi(t|\x_0)$ (rescaled by $R^2/D$) of the
reaction time on the spherical target of radius $R$ surrounded by an
outer reflecting concentric sphere of radius $L = 10R$, with $|\x_0| =
2R$.  We consider the truncated reactivity model with: {\bf (a)}
$\ell_1 = 0$ and several $\ell_2$, and {\bf (b)} several $\ell_1$ and
$\ell_2 = \infty$.  In all cases, $qR = 1$.}
\label{fig:models3}
\end{figure}

Figure \ref{fig:models3} illustrates more explicitly the effect of
truncated reactivity onto the reaction time distribution.  In the top
panel, we fix $\ell_1 = 0$ and consider several values of $\ell_2$,
i.e., the target is reactive at the beginning but becomes inert after
a prescribed number of encounters with the particle.  Expectedly, the
overall shape of the probability density $H_\psi(t|\x_0)$ does not
change much but it is progressively reduced as $\ell_2$ is getting
smaller.  This reduction results from the reduced normalization of
$\psi(\ell)$, which is equal to $1- \Psi(\infty) = 1- \exp(-q\ell_2)$.
For instance, for $q\ell_2 = 0.1$, the particle reacts only in $9.5\%$
of cases, explaining a tenfold decrease of $H_\psi(t|\x_0)$ in this
case (dash-dotted line).  In turn, the bottom panel of
Fig. \ref{fig:models3} presents the opposite situation when the target
is passive at the beginning and becomes reactive after a prescribed
number of encounters (i.e., we set $\ell_1 > 0$ and $\ell_2 =
\infty$).  The behavior of the probability density $H_\psi(t|\x_0)$ is
different.  When $\ell_1$ increases, more and more encounters with the
target at the beginning do not produce surface reaction, shifting
$H_\psi(t|\x_0)$ towards longer times.  Moreover, the shape of this
density progressively transforms from monomodal to bimodal that is
clearly seen at $q\ell_1 = 5$.  In fact, the first maximum of
$H_\psi(t|\x_0)$ is located around the time of the order of
$(|\x_0|-R)^2/(6D) \simeq 0.17 R^2/D$, which is controlled by the
distance to the target, $|\x_0|-R$.  This maximum corresponds to
direct trajectories from $\x_0$ to the target (see discussion in
\cite{Grebenkov18} for the conventional case).  As these rapid
trajectories fail to produce surface reaction at first encounters with
the inert target, the repeated returns to the target force the
particle to explore the confining domain.  As a result, the second
maximum emerges around the mean reaction time on partially reactive
target, which is of the order of $L^3(|\x_0|-R + 1/q)/(3DR^2)
\simeq 10^3 R^2/D$.  Such a bimodal shape of the probability density
of the reaction time is again a new feature induced by the considered
surface reaction mechanism.

\section{Comparison with bulk-mediated surface diffusion}
\label{sec:Chechkin}

In a series of works, Chechkin {\it et al.}
\cite{Chechkin09,Chechkin11,Chechkin12} and later Berezhkovskii {\it
et al.} \cite{Berezhkovskii15,Berezhkovskii17} investigated
bulk-mediated surface diffusion, which turns out to be complementary
to our approach.  In their model, a particle diffuses in the bulk with
diffusion coefficient $D_b$ until an encounter with the surface.
After hitting the surface, a particle can either resume bulk diffusion
or be adsorbed and start surface diffusion with diffusion coefficient
$D_s$.  After a random exponentially distributed time, the particle
desorbs from the surface and starts bulk diffusion again, until the
next adsorption.  This model enters the class of ``facilitated
diffusions'' \cite{Berg81} and was often employed in chemical physics
to describe reversible reactions \cite{Agmon90,Prustel13}, whereas
first-passage times of the associated intermittent process were
thoroughly investigated
\cite{Benichou10,Benichou11a,Rojo11,Rupprecht12a,Rupprecht12b,Rojo13}.

For cylindrical and planar domains, Chechkin {\it et al.} solved
coupled bulk and surface diffusion equations with exchange at the
surface and obtained an effective surface propagator $n_s(\s,t)$
characterizing the displacements of particles in the adsorbed state on
the surface.  For instance, in the case of a planar surface, this
propagator in the Laplace-Fourier domain reads
\cite{Chechkin12} (see Eq. (A6)):
\begin{equation} \label{eq:Chechkin1}
n_s(\k,p) = \frac{1}{p + D_s |\k|^2 + \tau_{\rm des}^{-1}[1 - W_{\rm bulk}(\k,p)]} \,,
\end{equation}
where $\k$ is the Fourier vector associated to $\s$, $\tau_{\rm des}$
is the mean desorption time, and $W_{\rm bulk}(\k,p)$ describes a
bulk-mediated excursion over a partially reactive surface.  This
function can be obtained by averaging our surface hopping propagator
$\Sigma_p(\s,\hat{\ell}|\s_0)$ over the stopping local time
$\hat{\ell}$ with exponential density $q e^{-q\ell}$.  In our
notations, it reads
\begin{align}  \nonumber
W_{\rm bulk}(\k,p) & = \F_{\k} \bigl\{ \E\{\Sigma_p(\s,\hat{\ell}|\s_0)\} \\
& = \frac{1}{1 + \frac{1}{q} \sqrt{|\k|^2 + p/D_b}} \,,
\end{align}
in agreement with Eq. (A5) from \cite{Chechkin12}, where the
coefficient $\mu D_b \tau_{\rm des} = a$ is identified with $1/q$, and
$\F_{\k}$ denotes the Fourier transform with respect to $\s-\s_0$.  In
the double limit $\tau_{\rm des} \to 0$ and $a \to 0$ (such that $\mu$
is kept constant), Eq. (\ref{eq:Chechkin1}) yields the effective
surface propagator on the plane:
\begin{equation} \label{eq:Chechkin2}
n_s(\k,p) = \frac{1}{p + D_s |\k|^2 + \mu D_b \sqrt{|\k|^2 + p/D_b}} \,.
\end{equation}
The properties of this propagator have been thoroughly investigated in
\cite{Chechkin12}.  We stress a significant difference between this
propagator and the surface hopping propagator $\Sigma_p(\s,\ell|\s_0)$
that we introduced.  For instance, we obtain in the Laplace-Fourier
domain:
\begin{equation}
\F_{\k} \{ \Sigma_p(\s,\ell|\s_0) \} = \exp\bigl(-\ell \sqrt{|\k|^2 + p/D_b}\bigr) ,
\end{equation}
which clearly differs from Eq. (\ref{eq:Chechkin2}), even in the case
of no surface diffusion ($D_s = 0$).

Berezhkovskii {\it et al.} analyzed coupled bulk-surface diffusion on
a flat surface by focusing on the cumulative residence times spent by
the particle in the two states \cite{Berezhkovskii15,Berezhkovskii17}.
They also derived the propagator in the Laplace-Fourier space, which
reads in our notations as
\begin{equation}
Q(\k,p) = \frac{1 + \frac{1}{q} \sqrt{|\k|^2 + p/D_b}}{p + D_s |\k|^2 + (p + D_s |\k|^2 + \tau^{-1}_{\rm des}) 
\frac{1}{q} \sqrt{|\k|^2 + \frac{p}{D_b}}} ,
\end{equation}
where $q = \kappa/D_b$.  This propagator could be directed deduced via
a renewal approach by summing the geometric series with bulk and
surface propagators:
\begin{equation}
Q(\k,p) = \frac{\tau_{\rm des} \, Q_s(\k,p)}{1 - Q_s(\k,p) W_{\rm bulk}(\k,p)} \,,
\end{equation}
with $Q_s(\k,p) = 1/(1 + \tau_{\rm des}(p + D_s |\k|^2))$ describing
surface diffusion.  We note that this approach differs from that by
Chechkin {\it et al.} as there is no double limit $q \to \infty$ and
$\tau_{\rm des} \to 0$.

In spite of this minor difference, both above approaches rely on
coupled bulk-surface diffusion, whereas our description focuses
exclusively on bulk excursions between encounters with the surface,
without absorption/desorption kinetics.  This description naturally
involves the boundary local time $\ell$ instead of physical time $t$
in the effective surface propagator.  As our description characterizes
reflections on the boundary before each successful binding, it may
provide deeper insights onto the intermittent dynamics and complement
the former approaches.

\section{Description of figure 1}
\label{sec:Fig1}

In this section, we provide some technical descriptions of Fig. 1 in
the main text.

The surface was generated by a freely available custom matlab routine
that produces a rough landscape with prescribed Hurst exponent $H$,
root-mean-square roughness $\sigma_S$, and number of pixels $M$
\cite{Kanafi19}.   We set $H = 0.95$ and $\sigma_S = 0.05$ to get mildly
rough surface of fractal dimension $3-H = 2.05$ and of size $1024
\times 1024$ pixels.  It was rescaled to get variations along $x$ and $y$
coordinates between $0$ and $1$.  Note that other types of surfaces
and other codes could be used for our illustrative purpose.

The random trajectory was simulated by our custom matlab routine that
generates independent three-dimensional Gaussian displacements with $D
= 1$ and timestep $\tau = 10^{-6}$ to have the standard deviation
$\sigma = \sqrt{2D\tau} \approx 1.41\cdot 10^{-3}$ along each
coordinate.  The initial point (red ball) was fixed at
$(0.5,0.5,Z(0.5,0.5)+h)$, where $Z(x,y)$ is the height of the surface
at $(x,y)$, and $h$ is a prescribed starting distance fixed at $h =
0.3$.  At each ($k$-th) step, it is checked whether the newly
generated position $(x_k,y_k,z_k)$ lies above the surface, i.e., $z_k
\geq Z(x_k,y_k)$.  In the opposite case, the $z$ coordinate is updated to
be $z_k = Z(x_k,y_k) + a$, with $a = 10^{-2}$.  This change implements
a reflection from the surface after an encounter (such encounter
points are shown by blue and yellow balls).  Reflecting boundary
conditions were also imposed at planes $x = 0$, $x = 1$, $y = 0$, and
$y = 1$ but we selected a trajectory that did not experience such
constraints.  A random trajectory with $10^5$ points was generated.

The surface points $(x,y,Z(x,y))$ were colored by using the function
\begin{equation}
\Sigma_0(x,y,\ell|x_1,y_1) = \frac{\ell}{2\pi[\ell^2 + (x - x_1)^2 + (y - y_1)^2]^{3/2}}  \,,
\end{equation}
where $(x_1,y_1,Z(x_1,y_1))$ is the first arrival position on the
surface, and $\ell = 0.3$ is a prescribed parameter.  This function is
maximal (dark red) near the first arrival point and then gradually
decreases as $(x,y)$ is getting farther from $(x_1,y_1)$ (dark blue).
If the surface was a plane, the function $\Sigma_0$ would be the
surface hopping propagator from $(x_1,y_1)$ to $(x,y)$ within the
local time $\ell$.  This is no more the case for a rough surface shown
in Fig. 1 but $\Sigma_0$ may still be considered as a lowest-order
crude approximation of the surface hopping propagator.  It aims just
to illustrate the basic idea behind this concept.


\end{document}